\PassOptionsToPackage{dvipsnames}{xcolor}
\documentclass[twocolumn]{aastex631}
\pdfoutput=1 
\usepackage[T1]{fontenc}
\usepackage{ae,aecompl}
\usepackage[utf8]{inputenc}
\usepackage[english]{babel}
\usepackage{amsmath,amstext}
\usepackage{apjfonts}
\usepackage{chngcntr}
\usepackage[figure,figure*]{hypcap}
\usepackage[hang]{footmisc}
\input{hyperlink-year-only-natbib-patch}
\allowdisplaybreaks
\graphicspath{{figures/}}

\newcommand*{\https}[1]{\href{https://#1}{\nolinkurl{#1}}}
\newcommand*{\http}[1]{\href{http://#1}{\nolinkurl{#1}}}

\defcitealias{Geha2017}{{Paper\,I}}
\defcitealias{Mao2021}{{Paper\,II}}
\defcitealias{saga3:Mao2024}{{Paper\,III}}
\defcitealias{saga4:Geha2024}{{Paper\,IV}}
\defcitealias{saga5:Wang2024}{{Paper\,V}}
\newcommand*{\paperone}{\citetalias{Geha2017}}
\newcommand*{\papertwo}{\citetalias{Mao2021}}
\newcommand*{\paperthree}{\citetalias{saga3:Mao2024}}
\newcommand*{\paperfive}{\citetalias{saga5:Wang2024}}

\newcommand*{\nsats}{378}

\newcommand*{\kms}{\ensuremath{\mathrm{km}\,\mathrm{s}^{-1}}}
\newcommand*{\msun}{\ensuremath{\text{M}_{\odot}}}
\newcommand*{\mstar}{\ensuremath{M_{\star}}}

\newcommand*{\EWHA}{\ensuremath{\text{EW}_{\text{H}_\alpha}}}
\newcommand*{\HA}{\ensuremath{\text{H}\alpha}}

\newcommand*{\satstable}{Table~C.3 of \paperthree{}}
\renewcommand*{\edit}[1]{#1}

\shorttitle{The SAGA Survey.\ IV.}
\shortauthors{Geha et al.}

\begin{document}
\title{The SAGA Survey.\ IV.\ The Star Formation Properties of 101 Satellite Systems around Milky Way-mass Galaxies}

\author[0000-0002-7007-9725]{Marla~Geha}
\affiliation{Department of Astronomy, Yale University, New Haven, CT 06520, USA}
\author[0000-0002-1200-0820]{Yao-Yuan~Mao}
\affiliation{Department of Physics and Astronomy, University of Utah, Salt Lake City, UT 84112, USA}
\author[0000-0003-2229-011X]{Risa~H.~Wechsler}
\affiliation{Kavli Institute for Particle Astrophysics and Cosmology and Department of Physics, Stanford University, Stanford, CA 94305, USA}
\affiliation{SLAC National Accelerator Laboratory, Menlo Park, CA 94025, USA}
\author[0000-0002-8320-2198]{Yasmeen~Asali}
\affiliation{Department of Astronomy, Yale University, New Haven, CT 06520, USA}
\author[0000-0002-0332-177X]{Erin~Kado-Fong}
\affiliation{Department of Physics and Yale Center for Astronomy \& Astrophysics, Yale University, New Haven, CT 06520, USA}
\author[0000-0002-3204-1742]{Nitya~Kallivayalil}
\affiliation{Department of Astronomy, University of Virginia, Charlottesville, VA 22904, USA}
\author[0000-0002-1182-3825]{Ethan~O.~Nadler}
\affiliation{Carnegie Observatories, 813 Santa Barbara Street, Pasadena, CA 91101, USA}
\affiliation{Department of Physics $\&$ Astronomy, University of Southern California, Los Angeles, CA, 90007, USA}
\author[0000-0002-9599-310X]{Erik~J.~Tollerud}
\affiliation{Space Telescope Science Institute, Baltimore, MD 21218, USA}
\author[0000-0001-6065-7483]{Benjamin~Weiner}
\affiliation{Department of Astronomy and Steward Observatory, University of Arizona, Tucson, AZ 85721, USA}
\author[0000-0002-4739-046X]{Mithi~A.~C.~de~los~Reyes}
\affiliation{Department of Physics and Astronomy, Amherst College, Amherst, MA 01002}
\author[0000-0001-8913-626X]{Yunchong~Wang}
\affiliation{Kavli Institute for Particle Astrophysics and Cosmology and Department of Physics, Stanford University, Stanford, CA 94305, USA}
\affiliation{SLAC National Accelerator Laboratory, Menlo Park, CA 94025, USA}
\author[0000-0002-5077-881X]{John~F.~Wu}
\affiliation{Space Telescope Science Institute, Baltimore, MD 21218, USA}
\affiliation{Center for Astrophysical Sciences, Johns Hopkins University, Baltimore, MD 21218, USA}

\correspondingauthor{Marla~Geha}
\email{marla.geha@yale.edu}

\begin{abstract}

We present the star-forming properties of \nsats\ satellite galaxies around 101 Milky Way analogs in the Satellites Around Galactic Analogs (SAGA) Survey, focusing on the environmental processes that suppress or quench star formation.  In the SAGA stellar mass range of $10^{6-10}\msun$,  we present quenched fractions, star-forming rates, gas-phase metallicities, and gas content.  The fraction of SAGA satellites that are quenched increases with  decreasing stellar mass and shows significant system-to-system scatter. SAGA satellite quenched fractions are highest in the central 100\,kpc of their hosts and decline out to the virial radius.  Splitting by specific star formation rate (sSFR), the least star-forming satellite quartile follows the radial trend of the quenched population.   The  median sSFR of star-forming satellites increases with decreasing stellar mass and is roughly constant with projected radius.  Star-forming SAGA satellites are consistent with the star formation rate--stellar mass relationship  determined in the Local Volume, while the median gas-phase metallicity is higher and median HI gas mass is lower at all stellar masses.  We investigate the dependence of the satellite quenched fraction on host properties. Quenched fractions are higher in systems with larger host halo mass, but this trend is only seen in the inner 100\,kpc; \edit{we do not see significant trends with host color or star formation rate}.  Our results suggest that lower mass satellites and satellites inside 100\,kpc are more efficiently quenched in a Milky Way-like environment, with these processes acting sufficiently slowly to preserve a population of star-forming satellites at all stellar masses and projected radii. 

\end{abstract}

\keywords{
\href{http://astrothesaurus.org/uat/416}{Dwarf galaxies (416)}, 
\href{http://astrothesaurus.org/uat/2040}{Quenched galaxies (2016)},
\href{http://astrothesaurus.org/uat/459}{Emission line galaxies (459)},
\href{http://astrothesaurus.org/uat/615}{Galaxy properties (615)},
\href{http://astrothesaurus.org/uat/2040}{Galaxy quenching (2040)}
}

\section{Introduction}
\label{sec:intro}

The star formation rate, metallicity, and  gas content of a galaxy provide an instantaneous snapshot from which its past evolution can
be inferred.   For low-mass galaxies ($\mstar < 10^{10}\msun$), these instantaneous properties are more stochastic than in more massive systems \citep[e.g.,][]{Weisz2012, Bauer2013,elbadry2016, emami2021}.   This stochasticity implies that low-mass galaxies are more susceptible to some combination of internal feedback and/or environmental conditions \citep[][]{peng2010A, samuel2022,pan2023}.    While the properties of isolated low-mass galaxies mainly provide insights into the internal processes regulating star formation \citep{geha2012,prole2021,Carleton2023},  the properties of low-mass satellite galaxies in orbit around a larger system are more complex, in that they reflect numerous processes including conditions in the larger host halo, the satellite's infall history, and its internal conditions prior to infall \citep{wetzel2013,buck:2019MNRAS.483.1314B,Christensen2023}

The satellites around the Milky Way have highly influenced our understanding of galaxy formation in low-mass satellite galaxies.   The Milky Way itself hosts just two satellites with active star formation: the LMC and SMC.  Both satellites have well-measured properties: bursty star formation histories \citep{Massana2022}, H$_2$ and HI gas reservoirs \citep{tumlinson2002,putman2021}, and emission in ultraviolet (UV) and H$\alpha$ due to hot young  stars heating the surrounding gas.  The orbits of the LMC and SMC \citep{Kallivayalil2013} suggest they are a galaxy pair on first approach into the Milky Way environment \citep{besla2007}. The remaining satellites within the Milky Way's virial radius (300\,kpc) are devoid of gas \citep{putman2021} and are considered ``quenched'', having no UV or H$\alpha$ emission since they ceased forming stars one or more billion years ago \citep{weisz2014}.   

Motivated by detailed observational data, tremendous theoretical effort has gone into understanding satellite systems in Milky Way-like environments \citep[e.g.,][]{Wetzel2016, Garrison-Kimmel2019:1806.04143, buck:2019MNRAS.483.1314B, 2022MNRAS.511.1544F, 2024ApJ...964..123J,brown2024}.   Gravity-only simulations predict thousands of low-mass dark matter subhalos around a Milky Way-mass central halo~\citep{2008MNRAS.391.1685S,2009MNRAS.398L..21S,2014MNRAS.438.2578G,Mao150302637,2016ApJ...818...10G,2023ApJ...945..159N}.  The distribution of luminous stars and gas within these subhalos depends sensitively on the treatment of galaxy formation processes (see \citealt{2022NatAs...6..897S} for a review). 
The stochastic nature of star formation in low-mass galaxies alone can lead to lower stellar masses, burstier feedback, and earlier star formation quenching \citep{2020MNRAS.492....8A}. Both cosmological~\citep{2021MNRAS.508.1652J} and zoom-in~\citep{2021ApJ...909..139A,vannest2023} hydrodynamic simulations of Milky Way-like environments suggest that environmental processes such as tidal or ram-pressure stripping act to quench star formation and remove gas from satellites \citep{Simpson17:1705.03018,Christensen2023}.  For example, \citet{2023MNRAS.522.5946E} demonstrated that ram pressure stripping is the dominant cause of quenching satellites around MW-mass hosts in the TNG50 simulations.   Satellites may also be `pre-processed' in a lower mass group environment prior to infall \citep{wetzel2013,samuel2022,2023MNRAS.519.4499P}.

A natural measure of success when modeling satellite systems in Milky Way-like environments is a direct comparison to observations.  However, the Milky Way's satellite population is a single snapshot of a single realization of a Milky Way-mass halo.   Increasing the number of satellite systems known around Milky Way-mass galaxies is critical yet challenging.   The primary challenge lies in discerning low mass galaxies from more massive, and far more numerous, background galaxies.   Out to a few Mpc, low mass galaxies can be identified via resolved individual stars \citep[e.g.,][]{Crnojevic2019, Doliva-Dolinsky2023,jones2023}.   These techniques have found low mass star-forming galaxies in the region just beyond the Milky Way \citep{mcquinn2023, Makarova2023} and satellites around a handful of massive host galaxies \citep{Nashimoto2022}.  Out to 20\,Mpc,  galaxies can be identified via surface brightness fluctuations (SBF) or criteria in galaxy properties such as surface brightness and radii \citep[][]{Danieli2018,Bhattacharyya2023}.   The largest effort to identify satellite galaxies via SBF is the Exploration of Local VolumE Satellites (ELVES) Survey \citep{Carlsten2019,Carlsten2022, greene2023}.   ELVES identified satellites down to $M_V\sim -9$ around two dozen hosts within 12 Mpc, 14 of which are similar to the Milky Way in terms of stellar mass and environment.

\begin{figure*}[htb!]
    \centering
    \includegraphics[width=1.0\textwidth,clip,trim=10 0 0 0]{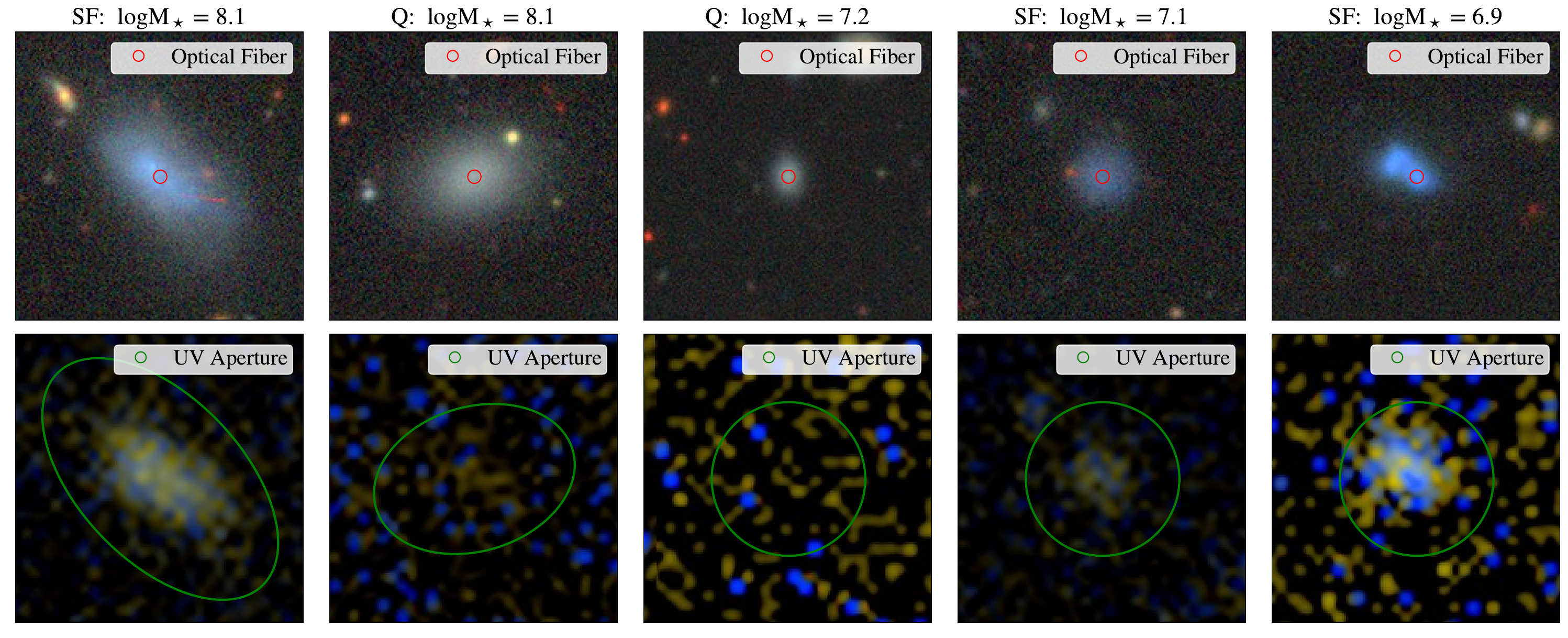}
    \caption{{\bf Top:}  Optical $grz$-composite  DECaLs images of example satellite galaxies in the SAGA Survey.   Red circles are the fiber coverage of the spectroscopic redshift measurement.   {\bf Bottom:} GALEX NUV/FUV-composite images for the same objects, with the aperture (green) used to measure NUV flux (the larger of $r_{\rm eff}$ or the $6"$ GALEX PSF).   The image title lists the satellite's stellar mass and whether it is classified as star-forming (SF) or quenched (Q). Satellites in the left two columns are in the Gold sample; the right three columns are Silver sample satellites.  All images are $0.75'$ on a side. }
    \label{fig:images}
\end{figure*}

The Satellites Around Galactic Analogs (SAGA) Survey\footnote{\https{sagasurvey.org}}
significantly expands on the rich observational data sets of low-mass galaxies in the MW and Local Universe.  SAGA is a targeted spectroscopic survey focused on Milky Way-mass hosts with distances between 25--40\,Mpc.  Initial SAGA results based on 8 Milky Way-analog host systems were presented in \citet[][\paperone{} hereafter]{Geha2017}.  The number of hosts increased to 36 in the SAGA results presented in \citet[][\papertwo{} hereafter]{Mao2021}.  In \citet[\paperthree{} hereafter]{saga3:Mao2024} , we present details for the completed survey of \nsats\ satellites around 101 Milky Way analogs.  \paperthree\ presents the full survey details and focuses on the luminosity function, radial distribution, satellite sizes, and quenched fractions. 

In this paper (Paper\,IV), we use  derived properties of SAGA satellites to explore the environmental processes that suppress or quench star formation.   In \S\,\ref{sec:data}, we briefly describe the SAGA satellite sample and provide details on how we measure H$\alpha$ and UV star-formation rates, emission line fluxes used for gas-phase metallicities and HI values from the literature.   In \S\,\ref{sec:quenched}, we explore the quenched fraction of satellites as a function of both stellar mass and projected radius.  In \S\ref{sec:star-forming-sats}, we focus on trends within the star-forming satellites.  In \S\ref{sec:gas}, we focus on observable properties of the gas in satellites, including a search for Active Galactic Nuclei, gas-phase metallicity and HI gas content.   In \S\,\ref{ssec_conformity}, we explore correlations between satellite quenched fractions and host properties.   In \S\,\ref{sec:discussion}, we consider what quantitative constraints the SAGA Survey can provide on quenching mechanisms, and then ask how the LMC/SMC themselves compare to satellites with similar luminosities in the SAGA Survey.   Finally, we summarize results in \S\,\ref{sec:summary}.  Further interpretation and comparison to theoretical models is given in \citet[\paperfive{} hereafter]{saga5:Wang2024}.

As in \paperthree, all distance-dependent parameters are calculated assuming $H_0 = 70$\,\kms\,Mpc$^{-1}$ and $\Omega_m = 0.27$.  Absolute magnitudes are $k$-corrected to $z=0$ using \citet{Chilingarian2010}. Magnitudes and colors are extinction corrected (denoted with a subscript `o,' e.g., $r_o$) using a combination of \citet{schlegel98} and \citet{Schlafly2011}.     We adopt a Kroupa initial mass function \citep[IMF;][]{kroupa2001} when computing  derived quantities and correct literature values to this IMF when making comparisons ($M_{*, {\rm Kroupa}} = 1.08M_{*, {\rm Chabrier}} = 0.66M_{*,{\rm Salpeter}}$).

\section{SAGA Satellite Data}
\label{sec:data}

The SAGA Survey Data Release 3 (DR3) spectroscopic targeting, satellite identification, and survey completeness are detailed in \paperthree\ and briefly summarized in \S\ref{ssec_saga}.   We provide measurement details for our estimates of optical line equivalent widths and fluxes in \S\ref{ssec_line_measure} and GALEX UV fluxes in \S\ref{ssec_nuv_measure}.   We provide our adopted star-formation rate prescriptions in \S\ref{ssec_sfr}.  In \S\ref{ssec_hi_lit}, we describe a literature search for HI flux measurements of SAGA satellites.

\begin{figure*}[htb!]
    \centering
    \includegraphics[width=1.0\textwidth,clip,trim=5 5 0 0]{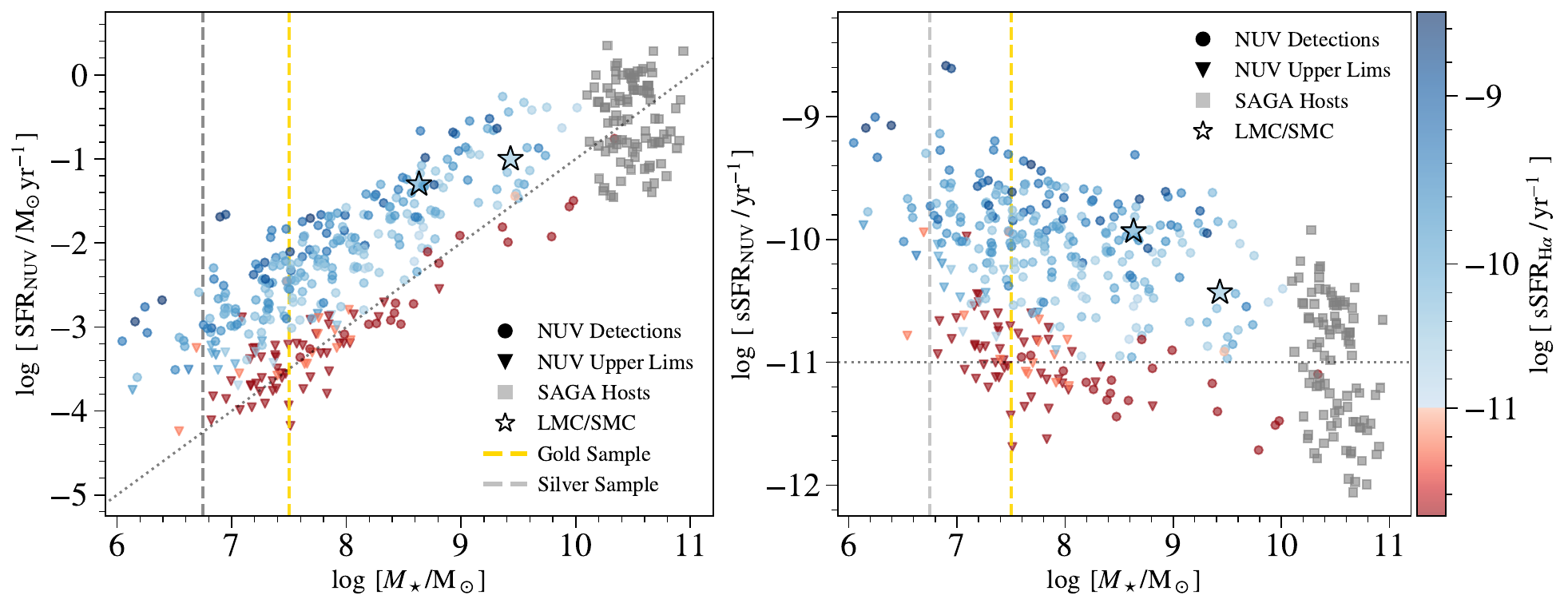}
    \caption{NUV-based star formation rates (SFR, left) and specific star formation rates (sSFR, right) versus stellar mass for all SAGA satellites with GALEX coverage (358 of 378 satellites).  Circles indicate NUV detections; triangles are NUV upper limits.   \edit{The GALEX NUV data are shallower than our H$\alpha$ observations (color-coding) which we use to detect satellites.}  We plot SAGA hosts (grey squares) and the LMC/SMC (stars, see \S\ref{ssec:MCs} for references).  In both panels, the dotted line indicates $\log [{\rm sSFR}_{\rm NUV}/{\rm yr}^{-1}]= -11$.  Vertical dashed lines indicate the stellar mass limits of our Gold ($M_{\star} > 10^{7.5}\,\msun$) and Silver ($10^{7.5} > M_{\star} > 10^{6.75}\,\msun$) samples.}
    \label{fig:SFMS_hosts}
\end{figure*}

\subsection{SAGA Data Release 3}\label{ssec_saga}

As in \paperthree{}, the SAGA DR3 sample consists of \nsats\ satellites around 101 hosts, representing a threefold increase in the number of hosts and satellites compared to our previous data release \citep{Mao2021}.  Milky Way-analog host galaxies are selected based on criteria in absolute magnitude ($-23 > M_K > -24.6$, a proxy for stellar mass) and environment over the distance range 25--40.75~Mpc (see \paperthree{}, \S2.2).  \edit{As described in \papertwo, the majority of our host distances are calculated based on the host's velocity, corrected for local peculiar velocity \citep{willick1997}.   When available, we use a redshift-independent distance measurement (26 of 101 hosts).}  As shown in Figure 3 of \paperthree{}, we identify five Local Group-like pairs among our 101 hosts (exactly two MW-mass hosts within 1 Mpc of each other), and an additional eight SAGA hosts that have precisely one MW-mass companion within 1 Mpc but the companion is not one of the 101 SAGA hosts.

We  identify potential satellite galaxy candidates based on photometric criteria using the photometric catalogs from DESI Legacy Imaging Surveys DR9 \citep[][]{Dey2019}. We obtain redshifts for candidate satellites via optical spectra from a variety of telescopes using both multi-fiber and single-slit spectrographs (\paperthree{}, \S3.2--3.3).   Satellites are defined as being within the projected virial radius of the host galaxy (300\,kpc, half a degree at 30\,Mpc) and having a projected velocity difference no greater than 275\,\kms~of the host redshift.  We exclude objects within 10\,kpc  of the host ($1'$ at 30\,Mpc) to avoid confusion.  We determine stellar mass for each satellite as described in \paperthree{} \S4.2, assuming $\log[\mstar/\msun] =1.254 + 1.098 (g - r)_o - 0.4 M_{r,o}$,  where color and absolute magnitude are corrected for Galactic extinction and assume a Kroupa IMF \citep{kroupa2001}.

The \nsats\ satellites in the SAGA DR3 sample span a stellar mass range $10^{6.0}- 10^{10.3}\,\msun$.  We split this sample into three regimes based on our estimated spectroscopic completeness (\paperthree{}, \S5.1).  The \textit{Gold} sample comprises satellites with a stellar mass greater than $10^{7.5}\,\msun$ and has high survey completeness for both quenched and star-forming satellites.   The \textit{Silver} sample comprises satellites with a stellar mass between $10^{6.75}$--$10^{7.5}\,\msun$.  The completeness correction for this sample remains reasonable for star-forming satellites, but is highly dependent on our completeness model for quenched satellites.  We include Silver satellites in some of the analyses below.  Finally, the \textit{Participation} sample comprises satellites with a stellar mass below $10^{6.75}\,\msun$.   While these satellites are shown in some plots,  the Participation sample is excluded from the analyses that follow due to the large uncertainty in their completeness correction.  Examples of satellites in the Gold and Silver samples are shown in Figure~\ref{fig:images}.   The stellar mass regions that define these samples are shown in Figure~\ref{fig:SFMS_hosts}.

\subsection{Measuring H$\alpha$ Equivalent Widths and Line Fluxes}\label{ssec_line_measure}

The SAGA satellites have optical spectra sufficient to detect H$\alpha$ in either emission or absorption as part of the survey design.  We measure the equivalent width of H$\alpha$ for all satellites.   For a subset of satellites where spectrophotometric flux calibration is possible, we measure line fluxes for several species.   In all cases, the optical spectrum covers only a portion of the galaxy;  the coverage fraction depends on both the size and stellar mass of the satellite.  Examples of fiber coverage are shown as red circles in the upper panels of Figure~\ref{fig:images}.

For every confirmed SAGA satellite, we first measure the equivalent width (EW) of H$\alpha$.    To measure EWs, we fit a single Gaussian line profile over a $20\mbox{\AA}$ region centered on the measured redshift.    We determine the H$\alpha$ EW by integrating the fitted profile.  We include the pixel variances in the fitting procedure and propagate these errors through to the fitted quantity.   These H$\alpha$  EW values are reported in \satstable{} and converted into star formation rates following the approach detailed in \S\ref{ssec_sfr}.

 While flux calibrated spectra were not a SAGA Survey requirement, we are able to reconstruct a flux calibration for many spectra as described in \citet{kadofong2024}.  For spectra from the AAT/2dF and MMT/Hectospec, we determine a relative flux calibration and use the galaxy's broad band magnitudes to infer an absolute flux scaling.   In these cases, we estimate line emission fluxes for species including H$\alpha$, H$\beta$, {[\ion{O}{3}]$\lambda$5007\AA{}},  [NII]$\lambda$6583\AA{}, and [SII]$\lambda$6717+6731 .  As described in \S3.1 of \citet{kadofong2024}, emission lines are fit simultaneously, assuming that all features are well-described by a Gaussian profile with the same width, but varying amplitude and velocity.  A Balmer absorption component is included as an ensemble of Gaussians with prescribed ratios of equivalent width for the progressively bluer Balmer lines.   We are able to measure emission line fluxes with S/N $>3$ in H$\alpha$ for 238 satellites, and S/N$>3$ in both H$\alpha$ and H$\beta$ for 166 satellites.  In this paper, line fluxes are used to determine an average internal reddening correction (\S\ref{ssec_sfr}), 
evaluate the ionization state of a galaxy (\S\ref{ssec:bpt}) and  estimate gas-phase metallicities (\S\ref{ssec_metallicity}).  These values are provided in \satstable.

\subsection{Measuring NUV Fluxes}\label{ssec_nuv_measure}

We estimate ultraviolet (UV) fluxes for SAGA satellites using data from the {\it Galaxy Evolution Explorer} \citep[GALEX;][]{Martin2005} which will be used as an additional indicator of star formation (\S\ref{ssec_sfr}).    We measure near-UV (NUV) fluxes directly from GALEX archival imaging.  This allows us to estimate meaningful upper limits in cases where GALEX is not deep enough to detect SAGA sources.  We do not re-measure FUV fluxes, as less than half of the SAGA satellite sample are detected in this band. 

To measure GALEX NUV flux, we first download a $5'\times5'$ cutout around each source from the DESI Legacy Sky Viewer\footnote{\https{legacysurvey.org}}.    To determine the background level and pixel variance we mask all sources using our optical catalog down to $r=21$. The background mask includes pixels inside 2 effective radii of catalog objects.   If the optical radius is less than the GALEX PSF ($6''$), we set the masked region to twice the PSF size.   We determine the image background and variance as the median and variance of the remaining background pixels.   We then perform aperture photometry on the original image, summing all flux inside an elliptical aperture determined from the optical photometry or within a circular $6"$ aperture if the optical radius is smaller than the GALEX PSF (Figure~\ref{fig:images}).   We flag non-detections assuming a conservative S/N < 4, separately flagging objects where no data is available.   GALEX imaging is available for 355 out of \nsats\ SAGA satellites (94\%).   We detect significant NUV flux for 276 out 355 satellites (78\%), and 257 out of 273 satellites identified as star-forming via H$\alpha$ emission (94\%).

We compare our NUV magnitudes to those measured by \citet{Karunakaran2021A} for our DR2 satellites, finding agreement to within 0.05\,mag once differences in the assumed foreground Galactic extinctions are taken into account.  Our NUV magnitudes and associated measurement errors are corrected for foreground extinction using \citet{Salim2016}, and reported in \satstable.  These values are converted into star formation rates in \S\,\ref{ssec_sfr}.

\subsection{NUV and H$\alpha$-based Star Formation Rates}\label{ssec_sfr}

We determine star formation rates (SFRs) using both H$\alpha$ EWs and our GALEX NUV flux measurements. H$\alpha$-based SFRs capture recent star-formation activity on the timescale of a few million years, whereas NUV traces longer timescales of order 100\,million years \citep[e.g.,][]{Lee2009,Broussard2019}.   We will directly compare these two indicators in \S\ref{ssec_sfs}. We do not use optical colors to infer SFRs as it is a far less quantitatively reliable indicator of star formation.

Our optical spectra cover only a portion of each galaxy (Figure~\ref{fig:images}).   To determine H$\alpha$-based SFRs, we therefore combine our H$\alpha$ EWs with the $r$-band absolute magnitude ($M_{r,o}$).  This provides an aperture corrected H$\alpha$ luminosity, assuming the galaxy has no strong color gradients \citep{Brinchmann2004}. \edit{The majority of satellites show at most mild color gradients (less than 0.1\,dex between the fiber and the rest of the galaxy).}  Following \citet{Bauer2013}, we first determine the H$\alpha$  luminosity as:
\begin{equation}
L_{\HA} = (\EWHA + \mathrm{EW}_{c}) 10^{-0.4 (M_{r,o} - 34.1)} \times\frac{3\times10^{18}[W]}{[6564.6(1+z)]^2},    
\end{equation}
\noindent
where $\EWHA$ are the measured values from \S\ref{ssec_line_measure}. $\mathrm{EW}_c$ corrects for stellar absorption, which we assume to be $2.5\mbox{\AA}$ for all satellites, a value consistent with our quenched satellite sample and the same as  \citet{Bauer2013}.  $M_{r,o}$ is corrected for Galactic foreground extinction and assumes the distance to the host.    

To correct $L_{\HA}$ for internal extinction, we determine an average correction  by directly measuring the line-flux ratio of $\HA/\mathrm{H}\beta$ (defined as the Balmer Decrement, BD) using the line fluxes measured above.   We measure a BD for less than half of the satellite sample.   We find an average value of BD = 3.25 with no dependence on stellar mass (see below for the corresponding $A_V$).    We apply this average internal correction to the full sample.  We assume an error of $\sigma_{\rm BD} = 0.05$ and propagate this, along with measurement errors, through to the SFRs. Again following \citet{Bauer2013}, we determine the $\HA$ SFR as:
\begin{equation}
{\rm SFR}_{\HA}\,[\msun\,{\rm yr}^{-1}]  = \frac{L_{\HA}}{1.27\times10^{34}}\times 0.66 \times \left( \frac{BD}{2.86}\right) ^{2.36},
\end{equation}
\noindent
where the first term is from \citet{Kennicutt1998A} and the factor of 0.66 converts this scaling from Salpeter to a Kroupa IMF using the value from \citet[][Figure 4]{Madau2014}.

We next convert our measured GALEX NUV magnitudes into star formation rates.   We first determine an internal extinction correction ($A_{\rm NUV}$) by assuming a \citet{Calzetti2000} reddening curve and $R_V =3.67$ determined by \citet{Battisti2016} for nearby star-bursting galaxies.  Using this reddening law, the average BD determined above (BD = 3.25) corresponds to an internal extinction value of $A_V = 0.4$ and $A_{NUV} = 0.9$.  We determine the NUV flux, $F_{\rm NUV}$, as:
\begin{equation}
F_{\rm NUV} = 10^{(m_{\rm NUV,o} - A_{\rm NUV} + 19.14)}
\end{equation}
\noindent
where $m_{\rm NUV,o}$ is our measured GALEX NUV magnitude, corrected for foreground Galactic extinction using \citet{Salim2016}. The NUV zeropoint of 19.14 corresponds to $F_{\rm NUV}$ in erg s$^{-1}$ cm$^{-2}$ \AA{}$^{-1}$. We can next determine the NUV luminosity as:
\begin{equation}
L_{\rm NUV} = {4\pi D^2}{F_{\rm NUV}}\frac{796}{3.826\times10^{33}},
\end{equation}
where D is the distance to the host in Mpc. We adopt an NUV filter width of $796$ \AA{} and compute NUV luminosities in units of $L_\odot$($=3.826\times10^{33}$ erg s$^{-1}$). Finally, we convert the corrected NUV luminosity into a SFR using Equation 3 of  \citet{2006ApJS..164...38I}:
\begin{equation}
{\rm SFR}_{\rm NUV}\,[\msun\,{\rm yr}^{-1}]  = \frac{L_{\rm NUV}}{2.14\times 10^9} \times 0.66,
\end{equation}
where the 0.66 factor is again correcting to a Kroupa IMF.   We note that the SFR$_{\HA}$ prescription uses the stellar spectral library of \citet{bruzual2003}, while  NUV is based on \citet{Lejune1997}.

For both SFR indicators, we define the specific star formation rate (sSFR) as the SFR divided by the stellar mass.   The error on the sSFR is determined from the quadrature sum of the SFR error and 0.2\,dex error on stellar mass.    SFR, sSFR and errors for both H$\alpha$ and NUV are reported in \satstable.   In Figure~\ref{fig:SFMS_hosts}, we plot SFR$_{\rm NUV}$ and sSFR$_{\rm NUV}$ versus stellar mass, color-coded by SFR$_{\HA}$.   We find broad agreement between the two SFR indicators and will compare these  in more detail in \S\ref{ssec_sfs}.

We note two SAGA satellites with very high specific star formation rates ($\log [{\rm sSFR}_{\rm NUV}/{\rm yr}^{-1}]= -8.6$) and stellar mass $\mstar\sim10^{6.9}\,\msun$, placing them well above the star-formation sequence seen in Figure~\ref{fig:SFMS_hosts}.    Both objects have several bright knots of star formation and lie far from their respective hosts (172 and 296\,kpc).  Neither show evidence for AGN in the form of broad Balmer emission or high ionization line-ratios (\S\ref{fig:BPT}).  Both galaxies are on the smaller side of the stellar mass-size relationship (\paperthree{}, Figure~12).    The optical and UV image for the second of these two systems is shown in the rightmost panels of Figure~\ref{fig:images}.  Both satellites are included in the analysis below.

\subsection{Literature HI observations}\label{ssec_hi_lit}

Ongoing star formation requires a gas reservoir.   To explore the atomic gas mass and gas fractions of our SAGA satellites (\S\,\ref{ssec_gas}), we have gathered literature HI measurements from single-dish 21\,cm radio surveys for 61 out of \nsats\ satellites.  In all cases, we recompute the HI mass from the published HI integrated flux measurements using the standard optically thin approximation \citep[Eq.\,9,\,][]{haynes1984}:  $M_{HI} = 2.356\times10^5\,D^2\,F_{HI}$  where $F_{HI}$ is the published integrated HI flux in ${\rm Jy\,km\,s}^{-1}$, and D is the distance to the host SAGA galaxy in Mpc.   To determine the total mass in atomic gas, we assume $M_{\rm gas} = 1.4 M_{HI}$, where the multiplicative factor takes into account the presence of helium and metals. 

To maintain homogeneity, we restrict our HI search to single-dish radio surveys.  We search for HI detections from three major HI surveys:  the HIPASS survey \citep[][64-meter Parkes Radio Telescope]{hipass2004}, the ALFALFA survey \citep[][300-meter Arecibo Telescope]{alfalfa2018}, and the FASHI survey \citep[][500-meter FAST Telescope]{Zhang2023}.     We match HI detections to SAGA satellites using the same spectrophotometric matching algorithm discussed in \paperthree{}, which includes both spatial and velocity matching.  We find a total of 61 detections:   2 matches with HIPASS, 26 matches with ALFALFA and 33 matches to FASHI.   We do not calculate upper limits for undetected SAGA sources within the survey footprints, although there are likely satellites whose upper limits would be informative.   We explore the 61 satellites with detected HI gas in \S\ref{ssec_gas}.

\section{Satellite Quenched Fractions}\label{sec:quenched}

The goal of this paper is to use the properties measured in the previous section to disentangle the environmental processes expected to suppress or quench star formation in SAGA satellites.   Ram-pressure stripping, tidal stripping, and suppressed gas accretion are all expected to contribute to quenching in MW-mass environments.   The relative strength and timescale of each process depends on an unknown combination of the host environment, and a satellite's orbital history, baryonic mass and total mass.  Given our observables, in the following sections we primarily investigate star-forming properties as a function of stellar mass and projected radius to address these questions.  

We will focus on the SFRs of SAGA satellites in \S\ref{sec:star-forming-sats}, but it is  useful to first simply classify galaxies as either star forming or quenched.   In this section, we describe our quenched definition (\S\ref{ssec_quenched}) and examine the fraction of quenched satellites as a function of stellar mass (\S\ref{ssec:fq}) and projected radius (\S\ref{ssec:fq_rad}).  We then compare the SAGA quenched fraction to a matched sample in the Local Volume (\S\ref{ssec:fq_ELVES}) from the ELVES survey \citep{Carlsten2022}.

\begin{figure}[t!]
    \centering
    \includegraphics[width=1.01\linewidth,clip,trim=5 5 0 0]{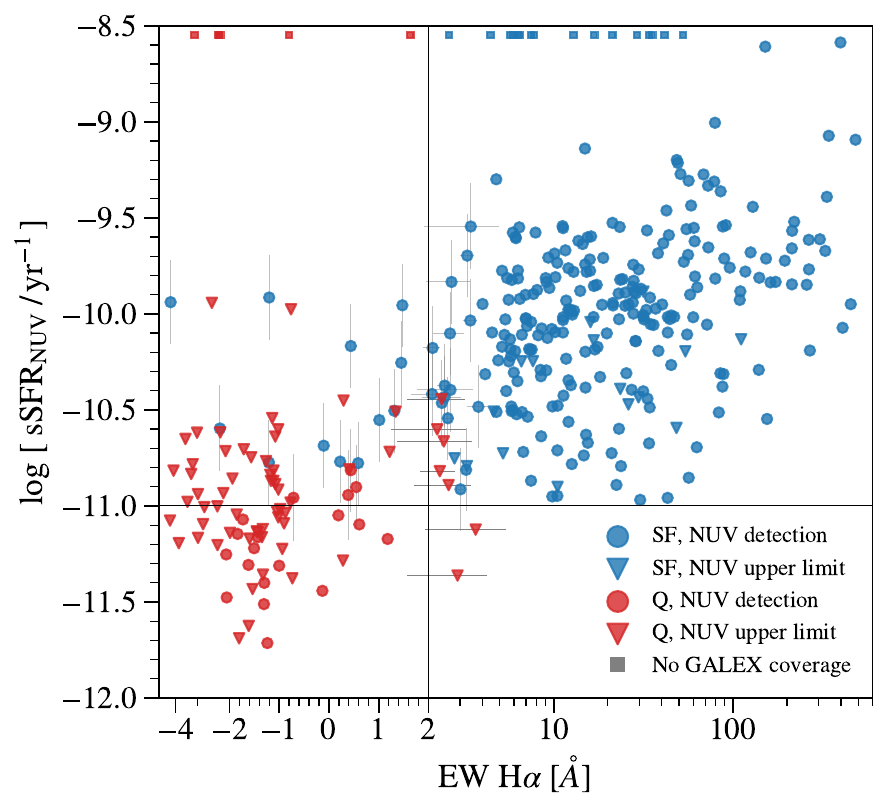}
    \caption{A SAGA satellite is defined as quenched (Q, red) or star-forming (SF, blue) based on combined criteria in H$\alpha$ EW and NUV specific star formation rate (sSFR$_{\rm NUV}$).  Circles indicate galaxies with detected flux in GALEX NUV, triangles indicate upper NUV limits.   Galaxies with no GALEX coverage are plotted as squares at the top of the plot.   Error bars are shown only for objects lying near the quenched boundaries (solid lines). Star-forming galaxies in the upper-left are those for which the central optical fiber clearly missed outer regions of star formation (see \S\ref{ssec_quenched}).} 
    \label{fig:quenched_def}
\end{figure}

\begin{figure*}[t!]
    \centering
    \includegraphics[width=1.01\textwidth,clip,trim=5 5 0 0]{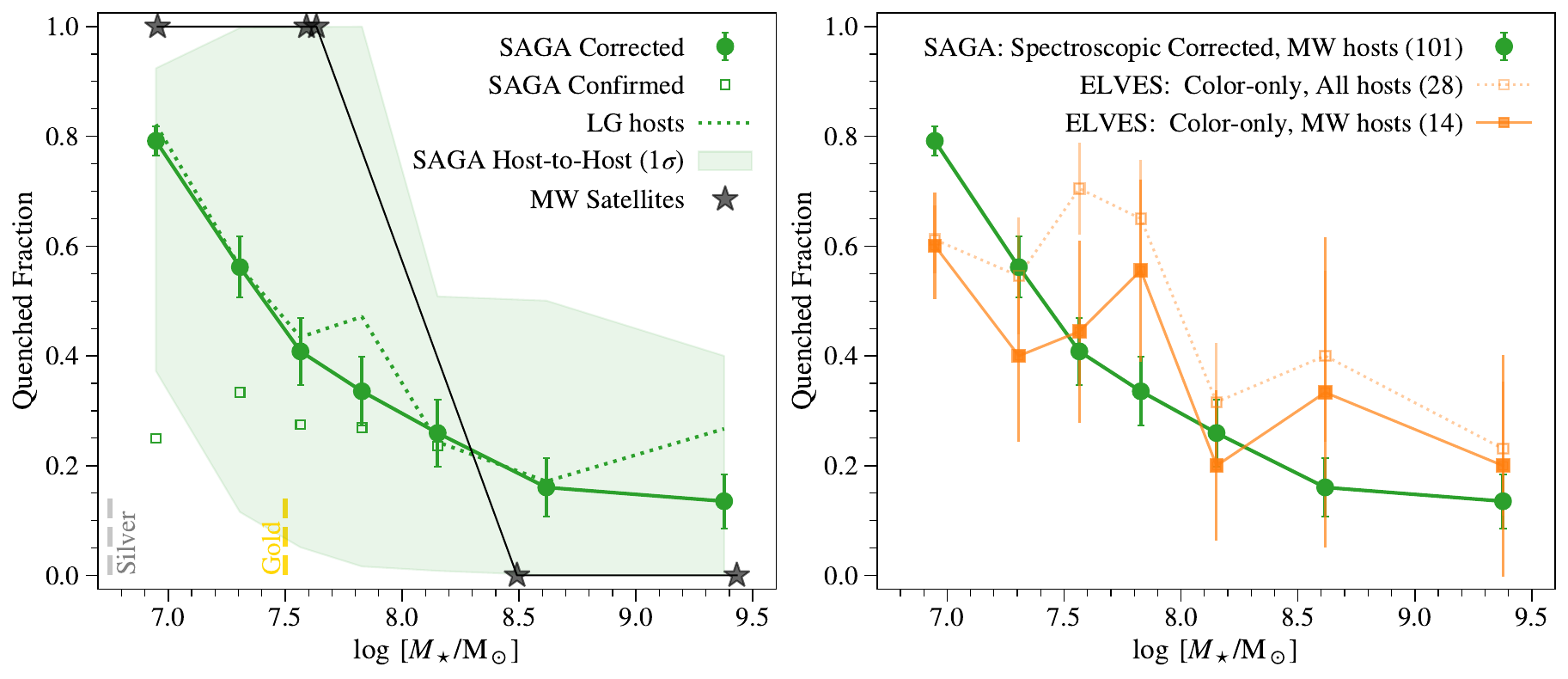}
    \caption{Quenched fractions as a function of stellar mass.   SAGA satellites, corrected for incompleteness, are shown in both panels as solid green circles with Poisson error bars.  {\bf Left:} We plot the uncorrected confirmed satellite quenched fraction (open green squares) which agree with the completeness corrected values for the Gold sample (gold bar), but increasingly deviate towards the Silver sample (grey bar).   The 1-$\sigma$ system-to-system scatter for the 101 individual SAGA systems is shown by the light green shaded region.   A subsample of 18 SAGA hosts in Local Group-like systems is shown as the dotted green line and is statistically indistinguishable from the full sample.   Black stars represent MW satellites.   {\bf Right:} We compare to the ELVES Survey \citep{Carlsten2022}, which classifies satellites using a color-only criteria.  We plot the full sample of 28 ELVES hosts (orange open squares) and a subsample of 14 ELVES hosts matching the SAGA MW criteria (solid orange squares, $-23>M_K>-24.6$). }
    \label{fig:fq}
\end{figure*}

\subsection{Quenching Definition}\label{ssec_quenched}

Our quenching definition assumes a galaxy is `quenched unless proven star-forming'.  We take this conservative approach due to our general finding that SAGA satellites are \emph{less} likely to be quenched than MW satellites.    We begin with the definition used in \papertwo{}, which considers a satellite as star-forming if $\EWHA> 2\,\mbox{\AA}$.   For SAGA DR3, we modify this criteria to include measurement errors, setting satellites as star-forming if $\HA$ is observed in emission with ($\EWHA - \sigma_{\EWHA})  > 2\,\mbox{\AA}$.   Seven galaxies are considered quenched due to the addition of this $\HA$-error criterion, seen as red symbols in the right two quadrants of Figure~\ref{fig:quenched_def}.   Our $\HA$ EW criteria is roughly equivalent to a sSFR$_{\HA} = -11.0\,{\rm yr}^{-1}$, however, we chose to implement this criteria in the observational space to maintain consistency with previous work.

The H$\alpha$ EWs alone are sufficient to classify the majority of satellites.  However, we introduce a second criteria based on sSFR$_{\rm NUV}$.   The majority of galaxies classified as star-forming by our $\HA$ EW criteria have $\log({\rm sSFR}_{\rm NUV}/{\rm yr}^{-1}) > -11.0$.   However,  a handful of our brighter satellites show a quenched core surrounded by a region of ongoing star formation which is missed in our optical fiber-based spectra (Figure~\ref{fig:images}). Thus, we also classify galaxies as star-forming if they have a specific NUV star formation rate: 
$ \log(\text{sSFR}_\text{NUV}/\text{yr}^{-1}) - \sigma_{\text{sSFR}, \mstar} > -11.0 $, 
where $\sigma_{\text{sSFR}, \mstar}$ combines the uncertainties in sSFR and stellar mass.%
\footnote{$\sigma^2_{\text{sSFR}, \mstar} = \sigma^2_{\left[\log(\text{sSFR}_\text{NUV}/\text{yr}^{-1})\right]}  + \sigma^2_{\left[\log(\mstar/\msun)\right]}$}
Out of \nsats\ satellites, 12 galaxies are classified as star-forming despite the lack of strong optical $\HA$ emission.  These objects are the blue circles in the upper left quadrant of Figure~\ref{fig:quenched_def}.  Visual inspection of these 12 galaxies confirms that the optical fiber missed the region of active star-formation in these systems.  The 22 SAGA satellites without GALEX coverage are shown as squares in Figure~\ref{fig:quenched_def}.

Our spectroscopic survey did not measure a redshift for every candidate satellite within our primary survey area.  As detailed in \paperthree{}, we determine a satellite probability for each candidate based on its magnitude, color, and surface brightness.   We use these probabilities to correct our satellites counts for spectroscopic incompleteness.
To correct our quenched fractions for spectroscopic incompleteness, we assess whether a given candidate is quenched using a color criterion.   In \paperthree{} Figure~12, we show that while there is no color criteria that perfectly separates the SAGA satellite population, the color-criteria defined by \citet{Carlsten2019}, translated into $g-$ and $r-$band, do an adequate job.   We apply these corrections to the quenched fractions. In the left panel of Figure~\ref{fig:fq}, the uncorrected (open green squares) and corrected (solid green circles) quenched fractions are similar for the  Gold sample, but increasingly deviate towards lower stellar mass (see also \paperthree{}, Figure 9).  We therefore restrict our analysis to Gold satellites when focused on quenched fractions (\S\ref{sec:quenched}), but combine Gold and Silver when focusing on the star-forming satellite population only (\S\ref{sec:star-forming-sats}).   Classifying our SAGA satellites using only color (ignoring our spectroscopic information) results in quenched fractions that are  over-estimated by 20--40\%, \edit{depending on the exact color criteria}.    This is consistent with the estimated bias in color-based quenched fractions in this stellar mass regime from SDSS \citep{geha2012} and ELVES \citep{Karunakaran2023}.

\subsection{Quenched Fractions: Trends with Stellar Mass}\label{ssec:fq}

We define the quenched fraction as the number of quenched satellites relative to the full population (quenched plus star-forming).  We plot the quenched fraction for SAGA satellites, corrected for spectroscopic incompleteness, as a function of stellar mass in Figure~\ref{fig:fq} (solid green circles with Poisson error bars).   The quenched fraction based on the raw confirmed satellites (open green squares) are similar to the corrected values for the Gold sample, but the magnitude of the incompleteness correction increases towards the Silver sample.   The shaded region in the left panel of Figure~\ref{fig:fq} shows the 1-$\sigma$ system-to-system scatter (16/84th percentiles) for our 101 SAGA hosts. The satellite quenched fraction based on our full sample of 101 hosts is similar to, but marginally higher than, what we reported in \papertwo{} based on 36 hosts. This  difference mostly comes from the improved incompleteness correction model, as discussed in \S4.5 of \paperthree{}.

The SAGA quenched fraction shown in Figure~\ref{fig:fq} increases steadily with decreasing stellar mass, from $15\pm5$\% above $10^{8.5}\msun$ (majority star-forming satellites) down to $79\pm3$\% near $10^{7}\msun$ (majority quenched).  In contrast,  the quenched fractions of isolated galaxies in the same stellar mass regime are nearly zero \citep{geha2012,prole2021,Carleton2023}.   The binned SAGA quenched fractions and Poisson errors are provided in Table~\ref{table:log_sm}.  \edit{To calculate the system-to-system scatter (green shaded region), we first obtain the quenched fraction for each host using consistent stellar mass bins and then determine the 16/84th percentiles in each stellar mass bin.} 

The overall quenched fraction of the Milky Way's satellite population follows the SAGA trend with stellar mass, but is more extreme (left panel of Figure~\ref{fig:fq}, black stars).  The Milky Way contains only two star-forming satellites (LMC/SMC); its remaining lower stellar mass satellites are quenched (i.e., the Milky Way's quenched fraction is zero above $10^{8.5}\,\msun$ and unity below).  
\edit{If we draw random samples from the average SAGA quenched fraction in the stellar mass range $10^{6.75-10}\msun$ using the stellar masses of the five most massive MW satellites, the Milky Way's quenched fraction is recovered in 16\% of the draws.  That is, the Milky Way’s quenched fraction is a 1-$\sigma$ outlier from the average SAGA quenched fraction distribution.} This can be seen visually in the left panel of Figure~\ref{fig:fq}:  the Milky Way's quenched fraction falls on the edge of the green shaded region  defined by the system-to-system scatter.   In this panel, we also plot the quenched fraction for the 18 hosts identified as Local Group-like (green dotted line).   We see no difference for Local Group-like hosts, compare to the full SAGA sample.   However, in \S\ref{ssec_conformity}, we describe subtle differences in quenched fractions for hosts with and without a nearby massive neighbor galaxy.

It is also interesting to consider the quenched fraction of satellites in our lowest stellar mass bin.  The SAGA Participation sample ($10^{6.0} <\mstar < 10^{6.75}\msun$) contains a number of low stellar mass star-forming satellites.   This sample is substantially incomplete, so the star-forming satellites at these stellar masses likely represent a much larger population, suggesting that the SAGA quenched fraction does not reach unity, even at these lower stellar masses.  Compared to the Milky Way which hosts no star-forming satellites below $10^{8.5}\,\msun$,  M31 contains one massive star-forming satellite, M\,33 ($10^{9.5}\,\msun$) and two lower mass star-forming systems: IC\,10 ($10^{7.9}\,\msun$) and LGS\,3 ($10^{6.0}\,\msun$).  Interestingly, these two lower mass satellites are at the very edge of the M31 virial volume.   The rarity of these lower mass star-forming satellites in the Local Group, and absence in the Milky Way, remains a puzzle.

\begin{figure}[t!]
    \centering
    \includegraphics[width=1.01\linewidth,clip,trim=5 5 0 0]{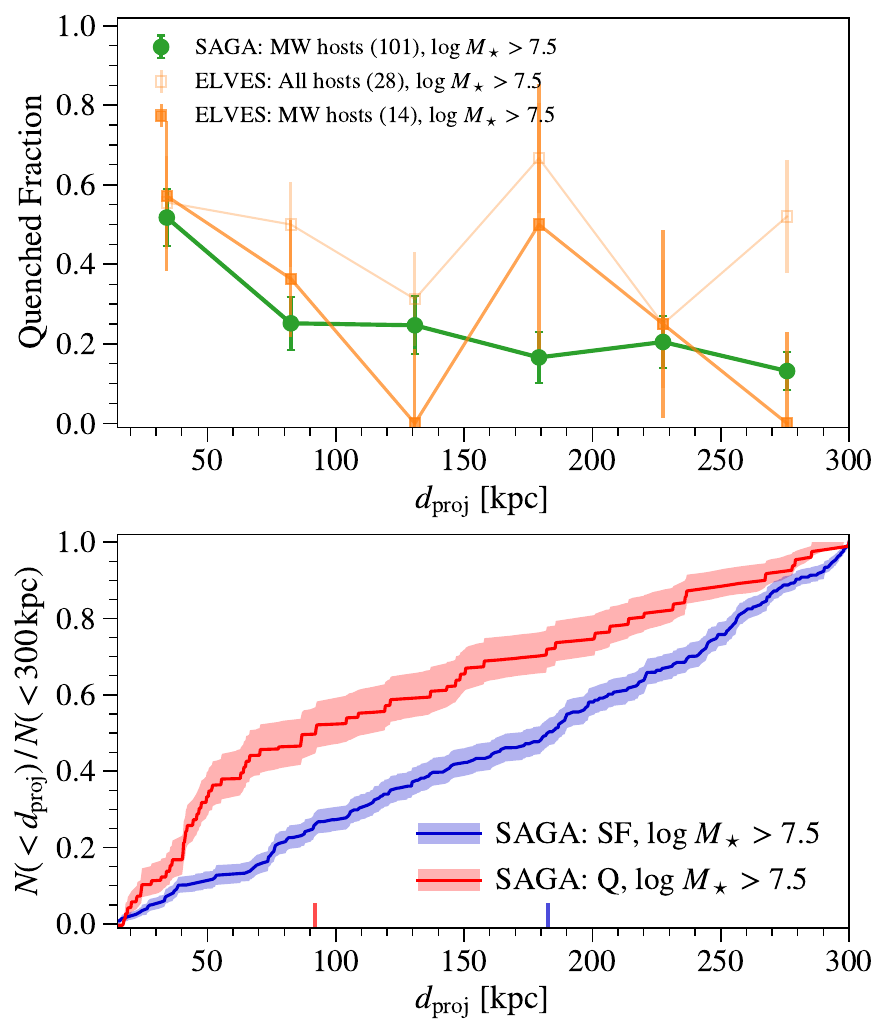}
    \caption{{\bf Top:} The SAGA quenched fraction (solid green) decreases with increasing projected distance from host (d$_{\rm proj}$).   We plot only the Gold satellites to minimize completeness corrections.   While the quenched fraction of the full ELVES sample (open orange squares) is higher at all radii, the matched sample largely agrees with SAGA (solid orange squares).   {\bf Bottom:}  Cumulative radial distribution function of SAGA quenched (red) versus star-forming satellites (blue) with projected distance from host. The shaded regions represent the contribution from incompleteness correction.   Vertical bars indicate the radius enclosing half of the satellites in each population.   Quenched galaxies are more radially concentrated than star-forming satellites at the same stellar mass. }
    \label{fig:fq_rad}
\end{figure}

\subsection{Quenched Fractions: Trends with Projected Radius}\label{ssec:fq_rad}

We next explore the projected radial distribution of quenched versus star-forming satellites.  In the bottom panel of Figure~\ref{fig:fq_rad}, we plot the cumulative radial distributions separately for quenched and star-forming satellites.  The shaded regions are computed by randomly resampling the incompleteness correction used to determine the cumulative distribution; the system-to-system variance is larger than this region.  We restrict this plot to Gold satellites ($\mstar > 10^{7.5}\msun$) where incompleteness corrections are small.   We explore the radial distribution within the star-forming satellite population in \S\ref{ssec_radial}.

In the bottom panel of Figure~\ref{fig:fq_rad}, the quenched Gold satellites (red) are more centrally concentrated than the Gold star-forming satellites (blue).   While quenched satellites with stellar mass greater than $10^{7.5}\,\msun$ exist throughout the virial volume of SAGA hosts, they are found closer to their host on average than are star-forming satellites at the same stellar mass.  Calculating the projected radius enclosing half of satellites, corrected for incompleteness, the half-radius for Gold quenched satellites is $92_{-26}^{+21}$\,kpc compared to $183_{-13}^{+7}$\,kpc for star-forming satellites.   As a result, the quenched fraction for the Gold sample (top panel of Figure~\ref{fig:fq_rad}, green solid line) is highest in the inner region, $52\pm 7\%$, decreasing quickly to $25\pm7\%$ by 80\,kpc and continuing a gentle decrease to $15\pm 7\%$ out to the virial radius (see Table~\ref{table:radial}).  Increased satellite quenched fractions in the inner region of a MW-mass halo has been seen in simulations \citep[e.g.,][]{1402.1498,Simpson17:1705.03018,2023MNRAS.522.5946E}; the physical scale at which this increase occurs constrains the quenching process.  For example, \cite{samuel2023} find that the effects of ram pressure stripping increases by several orders of magnitude from 100 to 10\,kpc in a MW-mass halo, a scale consistent with the increased quiescent fraction inside 100\,kpc in both SAGA and ELVES (see \S\ref{ssec:fq_ELVES}). 
 
While the quenched fraction of SAGA Gold satellites is highest in the inner region of SAGA systems, for the singular case of the Milky Way, the quenched fraction is currently {\em lowest} inside of 100\,kpc where the LMC/SMC reside. However, this is likely a temporary inversion explained by the recent infall of the LMC system into the Milky Way. Orbit calculations indicate that it is just past its first pericenter at $\sim 50$ kpc  \citep{Kallivayalil2013}, and will eventually merge with the MW in roughly $\sim 2$ Gyr \citep{cautun2019}.

\begin{figure*}[t!]
    \centering
    \includegraphics[width=1.\textwidth,clip,trim=0 0 0 0]{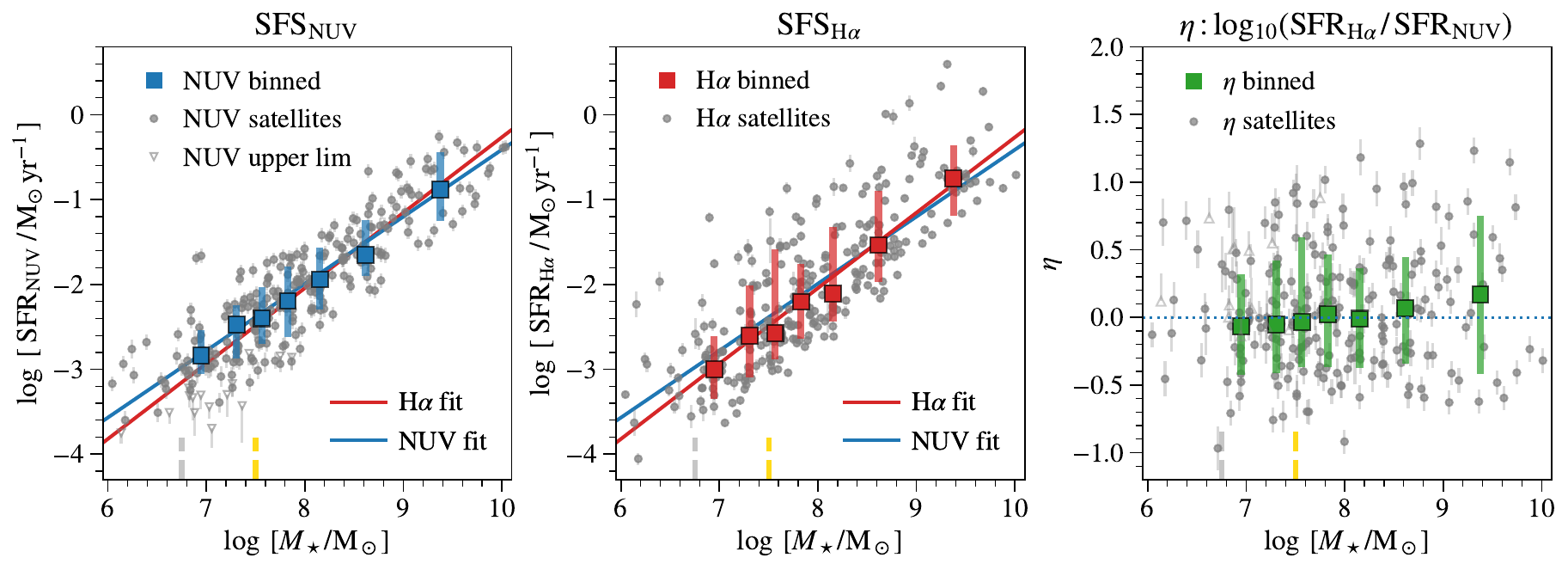}
    \caption{The star-formation--stellar mass sequence (SFS) for SAGA satellites based on our NUV SFRs ({\bf left}) and H$\alpha$ SFRs ({\bf middle}).   Quenched satellites have been removed; triangles indicated NUV upper limits but are not used to determine binned properties.  Binned mean and scatter (squares) for each sample are plotted, along with
    linear fits to the combined {\it Gold+Silver} sample for the NUV (blue) and H$\alpha$ (red).  {\bf Right:}  A direct comparison between SFR indicators, where $\eta$ is defined as the ratio of the H$\alpha$ to NUV SFR.   The mean binned value of $\eta$ is consistent with zero (green squares) indicating that satellite quenching timescale is longer than 100\,Myr.}
    \label{fig:SFMS_internal}
\end{figure*}

\subsection{Quenched Fractions:  SAGA vs. ELVES}\label{ssec:fq_ELVES}

We next compare the SAGA quenched fractions to a complementary set of satellites from the ELVES survey \citep{Carlsten2019,Carlsten2022,greene2023}.   ELVES surveyed 28 nearby host galaxy systems\footnote{\citet{Carlsten2022} states a full sample of 31 hosts.  We remove the MW, M31 and one host with no reported satellites to arrive at 28 hosts.} within 12\,Mpc using surface brightness fluctuations to characterize the distance of candidate satellites.  Compared to SAGA, the ELVES sample includes a larger range of host stellar mass ($-22.1 > M_K > -24.9$) and a less strict cut on environment.   ELVES classified a given satellite as quenched/star-forming based on a color-criteria in $g-$ and $i-$band.  We compare both to the full ELVES sample of 28 hosts and to a restricted sample of 14 hosts matched to the SAGA stellar mass and environment cuts.  See \paperthree{}, \S5.2 for the specific cuts made to match SAGA.

In the right panel of Figure~\ref{fig:fq}, we compare the SAGA quenched fraction (green circles) as a function of stellar mass to the full ELVES sample (open orange squares).   While both surveys show increasing quenched fractions with decreasing stellar mass, the full ELVES survey shows somewhat higher quenched fractions.  As noted at the end of \S\ref{ssec_quenched}, satellite selection based on color alone biases quenched fractions high.  Given the ELVES color selection, the true ELVES quenched fractions are likely lower by roughly 20\% in all bins \edit{(as estimated by comparing the quenched fraction of SAGA satellites using color-only selection, see \S\,\ref{ssec_quenched})}.   While this difference can explain the discrepancy with SAGA, we also plot the quenched fraction using only the 14 ELVES hosts which pass the SAGA host criteria (solid orange squares).    While the full ELVES sample shows higher quenched fractions, the MW-matched sample agrees within error with the larger SAGA sample.

In Figure~\ref{fig:fq_rad}, we continue the comparison to ELVES, exploring the trend in quenched fraction with projected distance to host.  In this plot, we restrict both samples to stellar mass $\mstar > 10^{7.5}\msun$.  We again compare both the full and MW-like ELVES samples.   The full ELVES sample appears flat with projected radius, while the MW-like ELVES hosts agree within errors to the SAGA result that quenched fractions are highest in the inner 100\,kpc.  These results agree with work by \citet{Karunakaran2023} who compared the SAGA DR2 sample (36 hosts) to a subsample of MW-like ELVES hosts using a UV-based sSFR to classify satellites.   In this case, \citet{Karunakaran2023} also found agreement between SAGA and ELVES quenched fractions as a function of stellar mass and that the quenched fractions decrease as a function of projected radius.

\section{Star-Forming Satellite Trends}\label{sec:star-forming-sats}

In this section we focus on trends within the SAGA satellite star-forming population, asking where and how star formation is suppressed in a MW-like environment.  For this section we expand our analysis to include both Gold and Silver satellites, as our completeness corrections for star-forming satellites remain minimal for $\mstar >10^{6.75}\msun$ \edit{(see left panel of Figure 9 in \paperthree{})}.  In \S\,\ref{ssec_sfs}, we  explore the star formation rate--stellar mass relationship,  comparing our NUV and H$\alpha$ SFRs to each other.  We then explore projected radial trends within the star forming satellite population in \S\ref{ssec_radial}.  Finally in \S\ref{ssec_sfs_lit}, we compare both quantities to galaxy samples outside MW-like environments.

\begin{figure*}[thbp]
    \centering
    \includegraphics[width=1.01\textwidth,clip,trim=0 5 0 0]{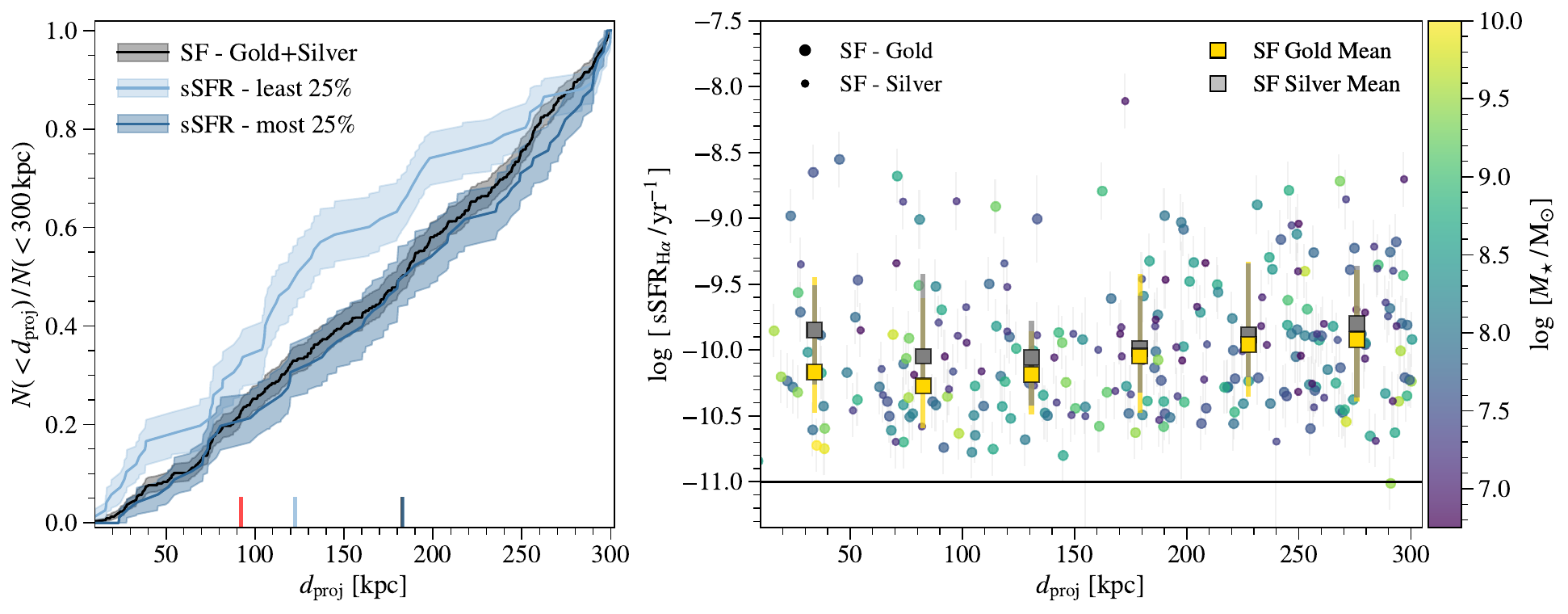}
    \caption{{\bf Left:} Cumulative number of star-forming satellites split into quartiles of sSFR$_{\HA}$ for the Gold+Silver sample.  The shaded regions represent the contribution from incompleteness correction.     The quartile with the lowest sSFR (lightest blue) follows the quenched population trend of being more concentrated towards the host.   We indicate the radii which encloses half the population as vertical bars for the full SF sample (black), the least/most sSFR quartiles (light/dark blue) and the quenched population (red).
    {\bf Right:} The sSFR$_{\HA}$ versus projected radius for star-forming satellites color-coded by stellar mass.   The median sSFR in bins of projected radius are shown for the Gold and Silver samples separately (squares).}
    \label{fig:sf_radial}
\end{figure*}

\subsection{The Star-Forming Sequence (SFS) for Satellites}\label{ssec_sfs}

Star-forming galaxies across all environments show a tight relationship between SFR and stellar mass, the so-called “Star-Forming Sequence” \citep[hereafter, SFS; e.g.,][]{Noeske2007, Popesso2023}.   This relationship can already be seen for the full SAGA satellite sample in the left panel of Figure~\ref{fig:SFMS_hosts}.   In this section, we remove quenched satellites, using the definition in \S\,\ref{ssec_quenched}, and compare the SFS for our NUV and H$\alpha$ SFRs.  We compare the SAGA SFS to literature in \S\,\ref{ssec_sfs_lit}.

In Figure~\ref{fig:SFMS_internal}, we show the SFS for SAGA satellites based on  SFR$_{\rm NUV}$ (left) and  SFR$_{\HA}$ (middle).  The SFR$_{\rm NUV}$ is tracing star-formation activity on 100\,Myr timescales, whereas SFR$_{\HA}$ traces timescales of order a few Myr \citep[e.g.,][]{Lee2009,Broussard2019,2008.08582}.   Qualitatively, these two SFS are similar, but the H$\alpha$-based SFS shows more scatter, as might be expected since it is averaged over shorter timescales.   More formally, we bin in stellar mass, plotting the median and 16/84th percentiles in each SFR and provide these values in Table~\ref{table:log_sm}.   We also fit a linear function to the combined Gold and Silver samples including errors in the measured SFRs.  In analyzing the NUV SFS relation, we use the star-forming galaxies in the upper two quadrants of Figure~\ref{fig:quenched_def} (266 galaxies); we exclude NUV upper limits but test that these do not significantly affect the results. In analyzing the H$\alpha$-based SFS, we use star-forming galaxies in the right two quadrants of Figure~\ref{fig:quenched_def} (273 galaxies).   

The median bins and linear fits for SFS in NUV and H$\alpha$ are nearly the same (compare red/blue lines in left and middle panel of Figure~\ref{fig:SFMS_internal}).  We estimate the RMS scatter around each SFS using the linear fits, finding that the scatter is smaller for NUV than for H$\alpha$.  The NUV SFS is slightly shallower than H$\alpha$, but well within the RMS scatter.   The additional H$\alpha$ scatter is biased towards higher SFR: the skewness of the H$\alpha$ distribution is 0.8 compared to 0.1 for the NUV distribution.  The higher scatter and skew in the H$\alpha$ SFRs can also be seen from the range of the 16/84th percentile bins in Table~\ref{table:log_sm}.

While we might expect H$\alpha$ SFRs to exhibit more scatter since it is averaged over a shorter time period, the skewness towards higher H$\alpha$ SFR (not seen in the NUV distribution) suggests that some fraction of satellites are currently under-going a short (<5\,Myr) burst of star formation.   We note that our $\HA$ SFRs are determined from a small region of the galaxy (Figure~\ref{fig:images}) and extrapolated assuming no color gradient.  This assumption likely introduces some scatter compared to the NUV, which is measured across the full galaxy.   

Our lowest stellar mass satellites have SFR between $10^{-3}$ to $10^{-4}\msun {\rm yr}^{-1}$.  In this regime, \citet{Lee2009} cautions that H$\alpha$ SFR begins to underestimate SFR compared to NUV due to the star-formation recipes themselves which are not calibrated to low mass galaxies.   This may also explain the small difference in slope between the two indicators.   It also suggests that future studies pushing towards lower stellar mass and lower absolute SFR will require renewed effort to calibrate SFR recipes in this regime.

In the right panel of Figure~\ref{fig:SFMS_internal}, we directly compare H$\alpha$ and NUV SFRs for individual galaxies.   Since our SFR indicators probe different timescales, their ratio provides some indication of star forming ``burstiness'' in a population \citep[e.g.,][]{2008.08582}.   Following \citet{Broussard2019}, we define the quantity $\eta$ as the ratio of the H$\alpha$ and NUV SFRs:  $\eta \equiv \log_{10}(\text{SFR}_{\HA}/\text{SFR}_{\rm NUV})$.  As defined, $\eta=0$ implies constant star formation over a 100 Myr timescale.

\begin{figure*}[htb]
    \centering
    \includegraphics[width=1.\textwidth,clip,trim=0 1 0 0]{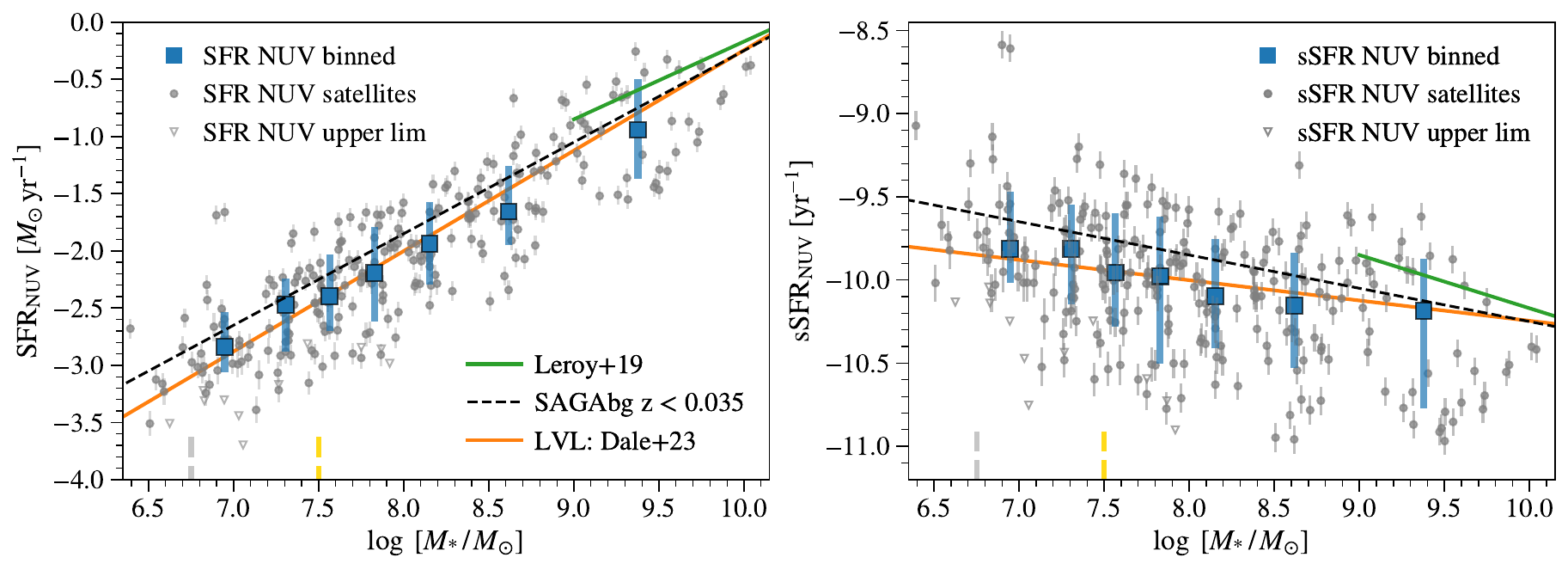}
    \caption{The SFR$_{\rm NUV}$ ({\bf left}) and  sSFR$_{\rm NUV}$ ({\bf right}) as a function of stellar mass.  SAGA star-forming satellites are plotted as grey circles, triangles indicated NUV upper limits.  Solid blue squares are binned medians, excluding upper limits.  The SAGA background (SAGAbg) galaxy sample ($z< 0.035$) is plotted as dashed black line.  
    \edit{The green line \citep{leroy2019} is fit at higher stellar mass and plotted down to stellar mass where it is well measured.  An extrapolation to lower stellar mass would disagree with SAGA samples.   The SAGA satellites and background sample agree well with the LVL sample (orange line) from \citet{dale2023}.  These samples are all fit in the same stellar mass regime.}}
    \label{fig:SF_MS}
\end{figure*}

The average $\eta$ for the SAGA satellite population is consistent with zero across our stellar mass regime (Figure~\ref{fig:SFMS_internal}, right panel).  The RMS scatter is $\sigma_{\eta} = 0.34$  and skewness is 0.3 towards positive $\eta$ values.   This suggests that the majority of star-forming satellites have constant SFR over 100\,Myr, while a small fraction of the population is undergoing a current burst.  We find no trend with $\eta$ and projected radius.  In particular, satellites with high $\eta$ are found at all projected radii.   A careful comparison to a matched isolated sample is required to determine whether this level of burstiness is consistent with isolated low mass galaxies or is enhanced in MW-like environments.

\subsection{SFR Trends with Projected Radius}\label{ssec_radial}

We have already explored the projected radial trends of quenched versus star-forming SAGA satellites in \S\ref{ssec:fq}, finding  that quenched Gold satellites are  more radially concentrated as compared to the  star-forming population (Figure~\ref{fig:fq_rad}).  Here we focus on radial trends within the star-forming satellite population itself.  

In the left panel of Figure~\ref{fig:sf_radial},  we plot the cumulative distribution of star-forming satellites with projected radius.  We split this population into quartiles based on the specific star formation rate  ($\text{sSFR}_{\HA}$) combining Gold and Silver satellites ($\mstar > 10^{6.75}M_{\sun}$).   This plot is qualitatively the same if we split using sSFR$_{\rm NUV}$ or use the Gold sample alone.  We find that the least star-forming quartile (light blue) is significantly more centrally concentrated than each of the other 3 quartiles individually or combined.  Computing the projected radius enclosing half of a given population, the least star-forming quartile half-radius  is $120_{-10}^{+12}$\,kpc.   In comparison, the next 3 quartile half-radii are $184_{-13}^{+14}, 206_{-16}^{+17}$ and $185_{-14}^{+17}$\,kpc.   The combined half-radius for the full star-forming population is $180_{-7}^{+10}$\,kpc.    Thus, the radial distribution of the three most star-forming quartiles are similar, while the least star-forming satellites are distributed more like quenched satellites whose half-radius is 92\,kpc (computed in \S\ref{ssec:fq_rad}).   This suggests that the least star-forming satellites are in the process of quenching and will soon join the quenched satellite population.   The observation that a quarter of star-forming satellites appear to be in the process of quenching provides constraints on the timescale and relevant physical processes for environmental quenching, however, further quantifying this timescale requires a full modeling framework (\S\ref{ssec:MCs} and \paperfive{}).

In the right panel of Figure~\ref{fig:sf_radial}, we plot the sSFR$_{\HA}$ of individual galaxies versus projected radial distance, color-coded by stellar mass. SFR$_{\HA}$  traces shorter timescales as compared to NUV and might reflect subtle radial trends not observed in the cumulative distributions.   Instead, we find that satellite sSFR$_{\HA}$ are evenly distributed with projected radius, including satellites with the highest sSFR$_{\HA}$.   The plot with sSFR$_{\rm NUV}$ is qualitatively similar.   We plot the median and 16/84th percentiles binned in projects radius for the Gold and Silver star-forming satellites separately (squares, right panel Figure~\ref{fig:sf_radial}).   Lower stellar mass satellites have slightly higher sSFR at all radii (see \S\ref{ssec_sfs_lit} and Figure~\ref{ssec_sfs_lit}).   \edit{The median sSFR is largely constant with projected radius.   The innermost sSFR bin shows slightly higher sSFRs, despite our previous observation that satellites with the lowest sSFRs are concentrated in the inner regions.}   We provide the binned median values for the sSFR versus project radius in  Table~\ref{table:radial}.

\begin{figure*}[htb]
    \centering
    \includegraphics[width=1.0\textwidth,clip,trim=0 0 0 0]{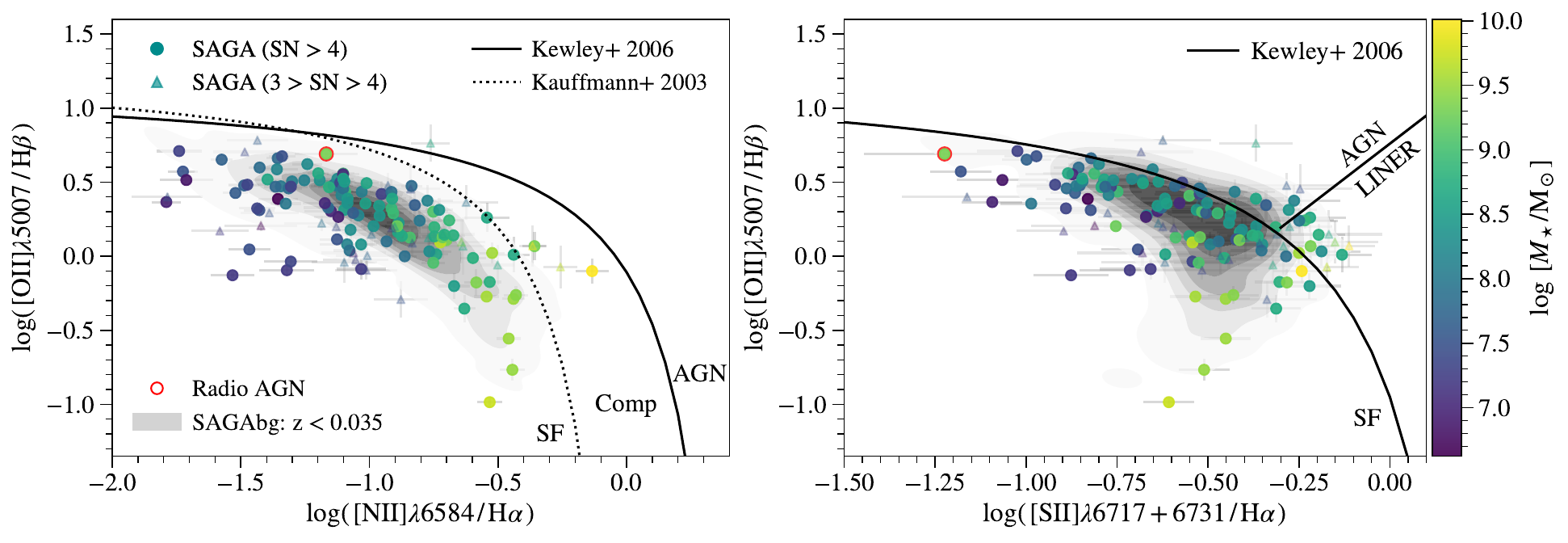}
    \caption{The BPT ({\bf left}) and [SII]-BPT ({\bf right}) diagram for SAGA satellites color-coded by stellar mass.   The SAGA background sample ($z < 0.035$ and $M_{\star}<10^{10}\msun$) is plotted as grey contour-scatter.   We restrict both samples to galaxies with $S/N > 4$ in all lines, but also plot satellites with a lower threshold $3 > S/N > 4$ for comparison (triangles).   Regions of the BPT where emission is dominated by star formation (SF), AGN, or composite (Comp) are labeled.   No high S/N SAGA satellites fall within the AGN region as defined by \citet[][solid line]{kewley2006} or \citet[][dotted line]{Kauffmann2003}.  The single SAGA satellite AGN confirmed via radio continuum does not fall in the optical AGN region (red circle).   The SAGA satellite population roughly follow the background distribution. }
    \label{fig:BPT}
\end{figure*}

\subsection{Comparing SAGA SFR Trends to the Literature}\label{ssec_sfs_lit}

Having established that, for star-forming SAGA satellites, the sSFRs do not change significantly with projected radius, we next revisit the stellar mass--SFR relation, comparing the SFS of SAGA satellites within the virial radius to galaxies in a wider range of environments. The zero point of the SFS increases with redshift \citep[e.g.,][]{Noeske2007, leja2022}, therefore, we restrict our comparison to redshift samples with $z < 0.035$.   This corresponds to the lowest redshift bin in \citet{kadofong2024} that shows no evidence of SFS evolution.   In Figure~\ref{fig:SF_MS}, we plot the NUV-based SFS and sSFR for SAGA satellites.  We focus on the NUV SFRs, which have similar slope/zeropoint to H$\alpha$ for SAGA satellites (\S\ref{ssec_sfs}), but lower scatter.   

We first compare the SAGA satellite SFS to the SAGA background sample (dashed line in Figure~\ref{fig:SF_MS}).  The SAGA background sample contains galaxies within the SAGA observing footprint which are not associated with the host (see \paperthree{} \S4.4).  The background sample is observed and processed using the same pipelines as SAGA satellites, including the same NUV SFR prescriptions.  The SAGA satellite median (blue squares) and the background population (dashed) are similar to within errors.   We find that the background sample has a smaller rms scatter compared to the satellites (0.2 vs.~0.3), which remains the same if we restrict the comparison to stellar mass regime $10^7 < \msun < 10^9$ where the samples overlap the most.   The higher scatter in the satellites is driven by the lowest sSFR galaxies which are present, but in lower relative numbers, in the background population.

We next compare the SAGA satellite SFS fits in the literature.  We compare to published fits from the Local Volume (LVL) sample of \citet{dale2023} and the z0MGS sample from \citep{leroy2019}.  In both cases, we correct the published SFS fit to a Kroupa IMF.     The star-forming SAGA satellites appear to fall along the LVL SFS (orange), \edit {but would systematically fall below an extrapolation of the z0MGS relation (green)}.  The published z0MGS  relationship is determined for galaxies in a higher stellar mass and slightly higher redshift regime, however, we have also refitted this sample using the same redshift constraints and find a similar fit.   The LVL relationship is better matched in stellar mass to the SAGA satellites but represents a wider range of environments.  The slope and zeropoint of the LVL is the same within errors to the SAGA satellites, but again shows a lower scatter around this fit of 0.2.   We conclude that the slope of the SAGA satellite SFS is consistent with other samples, but shows additional scatter driven by the lowest sSFR satellites. These satellites are the same satellites found to be more radially concentrated (\S\ref{ssec:fq_rad}) and are likely in the process of quenching star formation due to environmental processes more active in MW-like environments.

\section{Gas Properties of SAGA Satellites}\label{sec:gas}

We next focus on the observable properties of gas in SAGA satellite galaxies.  The ionization state and metallicity of gas in a galaxy can be assessed via optical emission lines, while the amount of atomic gas is estimated via radio HI measurements.  In this section we search for Active Galactic Nuclei (AGN; \S\ref{ssec:bpt}), trends with gas phase metallicity (\S\,\ref{ssec_metallicity}) and HI gas mass (\S\,\ref{ssec_gas}).

\subsection{A Search for AGN and the BPT Diagram}\label{ssec:bpt}

In massive galaxies ($\mstar > 10^{10}\msun$), AGN feedback is a primary process for quenching galaxies \citep[e.g.,][]{Kauffmann2003, Combes2017}.   In lower stellar mass galaxies, it is harder to separate AGN from star-formation signatures \citep{reines2022}. As a result, the usual AGN detection methods are less efficient and require complementary methods to build a complete AGN census \citep{Baldassare2018,Polimera2022}. Thus, despite AGN detections in numerous low mass galaxies, the influence of AGN on galactic quenching remains unclear \citep[e.g.,][]{dickey2019, Davis2022, Schutte+2022} and may even enhance star formation in low mass galaxies \citep{Cristello2024}. 

We search for multi-wavelength evidence of AGN starting with the X-ray archives from Chandra \citep{evans2022} and the eRASS1 data release from eROSITA \citep{erass1}.  Roughly 200 satellites have coverage in one or both X-ray archives, however, we find no conclusive matches.   At mid-infrared wavelengths,  the majority of SAGA satellites (292) are detected in WISE W1 and W2 bands, however, no SAGA satellites pass the  \citet{stern2012} AGN criteria of  (W1$-$W2$) > 0.8$. 
AGN can also be detected as flat-spectrum central radio sources \citep{Davis2022}.  We perform a visual search of the Very Large Array Sky Survey \citep[VLASS; ][]{lacy2020} using the reduced images available in the Legacy survey \citep{Dey2019}.  One satellite (OBJID: 902799740000002207) hosts a strong central radio continuum source out of 325 SAGA satellites with VLASS imaging.  We highlight this radio source below.

We next search for AGN using optical emission-line diagnostics.   Using line fluxes measured from our SAGA spectra (\S\ref{ssec_line_measure}), we construct the [OIII]$\lambda$5007\AA{}/H$\beta$ versus [NII]$\lambda$6584\AA{}/H$\alpha$ diagnostic diagram \citep[the BPT diagram;][]{Baldwin1981} and the [SII]-BPT based on [SII]$\lambda$6717+6731/H$\alpha$ \citep{VO87_bpt}.   In Figure~\ref{fig:BPT}, we plot these two diagnostics for both SAGA satellites and the SAGA background sample described in \S\ref{ssec_sfs_lit} ($z < 0.035$, $\mstar < 10^{10}\msun$).  In both cases, we set a S/N threshold above 4 in all line fluxes, however, to increase the satellite sample, we also show satellites using a lower S/N threshold (3 < S/N < 4, triangles in Figure~\ref{fig:BPT}).  

No SAGA satellites with well-measured line fluxes fall into the AGN-demarcation regions defined by \citet{kewley2006} and \citet{Kauffmann2003} in Figure~\ref{fig:BPT}.   This includes the radio-detected AGN noted above (red circle, Figure~\ref{fig:BPT}).  One SAGA satellite  with lower S/N falls in both the BPT and [SII]-BPT AGN region (OBJID:  902881160000004697), and two additional lower S/N satellites fall in the [SII]-BPT AGN region.    These satellites do not show other evidence for AGN.  Given their low S/N, we do not consider these secure detections.   No SAGA satellite shows evidence for broad Balmer lines.   Thus, we conclude the optical spectra of our SAGA satellites are dominated by emission from star formation.  In \S\ref{ssec_metallicity}, we use the optical spectra to determine gas-phase metallicity without removing any satellites.    

The SAGA satellites follow the overall distribution of the SAGA background population in the BPT diagram, however, we note an excess of satellites falling in the LINER region of the [SII]-BPT (high values of [SII]$\lambda$6717+6731/H$\alpha$) as compared to the background.  All satellites in this region ([SII]$\lambda$6717+6731/H$\alpha > -0.25$) also have gas-phase metallicities in excess of the main SAGA relationship which we will discuss in \S\ref{ssec_metallicity}.

In the SAGA stellar mass regime, AGN occupation fraction estimates in the literature range \edit{from 0.3\% to over 20\%} \citep{Polimera2022,reines2022}.   
Combining the multiple AGN indicators above, we find only one SAGA satellite with secure evidence of hosting an AGN.  From this single object, we place a lower limit of 0.3\% on the AGN occupation fraction in SAGA satellites.  \edit{However, deeper observations and/or alternative AGN selection methods which are more sensitive to AGN in the low mass galaxies \citep[e.g., photometric variability;][]{Baldassare2018} may uncover additional AGN in this population.}

\begin{figure}[t!]
    \centering
    \includegraphics[width=1.0\columnwidth,clip,trim=5 0 5 0]{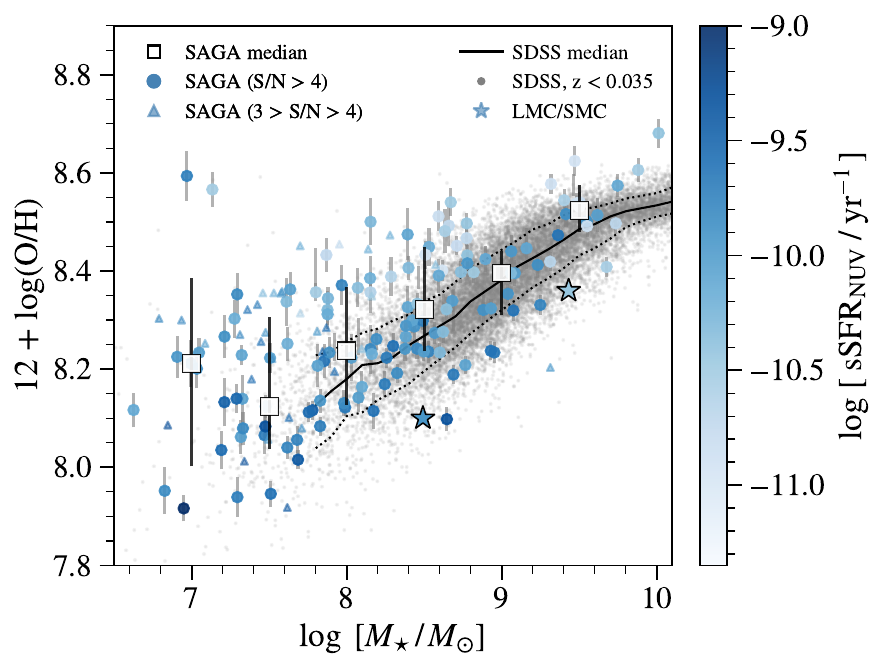}
    \caption{Gas-phase metallicity, 12+log(O/H), for SAGA satellites color-coded by sSFR$_{\rm NUV}$.   White squares are binned SAGA medians and 16/84th percentile scatter.    We compare to SDSS DR8 (grey) for galaxies with $z< 0.035$.  All metallicities have been  derived using the N2 index and \citet{marino2013} calibration.   Only individual satellites with line fluxes with S/N > 4 (circles) are included in the medians; SAGA satellites with line fluxes 3 > SN > 4 (triangles) are shown for comparison.   We plot the LMC/SMC metallicities as stars (see \S\ref{ssec:MCs} for references).    SAGA satellites lie at slightly higher gas-phase metallicities as compared to the SDSS sample. Satellites with higher sSFRs lie at lower gas-phase metallicities for a given stellar mass.} 
    \label{fig:metallicity}
\end{figure}

\subsection{Gas-Phase Metallicity}\label{ssec_metallicity}

A galaxy's metallicity is positively correlated with its total stellar mass, as metals are formed and recycled into the interstellar medium over generations of star formation \citep[e.g.,][]{kewley2008, berg2012,telford2016}.   The detailed relationship between these quantities informs models of star formation feedback and the inflow/outflow of galactic gas \citep[e.g.,][]{torrey2019,buck:2019MNRAS.483.1314B}.  Gas-phase metallicity is often specified by the oxygen abundance, 12 + log(O/H), as a proxy for the global gas-phase metallicity.  \citet{tremonti2004} first showed that below $\mstar \sim 10^{10}\msun$, gas-phase metallicity decreases roughly linearly with the log of stellar mass.

Given the wavelength coverage and signal-to-noise of our spectra, we choose to measure gas-phase metallicity using the strong line diagnostic N2 \citep{Pettini2004,kewley2008}:
\begin{equation}
\mathrm{N}2 = \log \left(\frac{[\mathrm{NII}]\lambda6584}{\HA}\right)
\label{eq:n2}
\end{equation}
\noindent
We convert this N2 index into gas-phase metallicity using the calibration of \citet[][Eq.~4]{marino2013}:  
\begin{equation}
12 + \log ({\rm O/H}) = 8.743 + 0.462 \times \mathrm{N2}
\label{eq:metallicity}
\end{equation}
\noindent
While this is an empirical calibration, we find a small mean offset of $\Delta (12+\log({\rm O/H}))=0.04$ for the small number of galaxies (14 satellites, 40 SAGAbg galaxies at $z<0.05$) in which direct auroral line metallicities can be inferred \citep{kadofong2024}. We have additionally tested the O3N2 index \citep{marino2013} and find no qualitative difference.    We focus on a relative comparison between SAGA satellites, the SAGA background sample and a low redshift sample from SDSS.   For each sample, we re-compute all gas-phase metallicities using the N2 index and \citet{marino2013}.   This is critical as the difference between samples is smaller than the difference between calibrations:  the absolute metallicity scaling  between calibrations can differ by up to 0.7\,dex \citep{kewley2008}.    We provide our measured line fluxes in \satstable, so that metallicity can be recomputed if a different calibration is desired.

\begin{figure*}[htb!]
    \centering
    \includegraphics[width=1.02\textwidth,clip,trim=1 0 0 0]{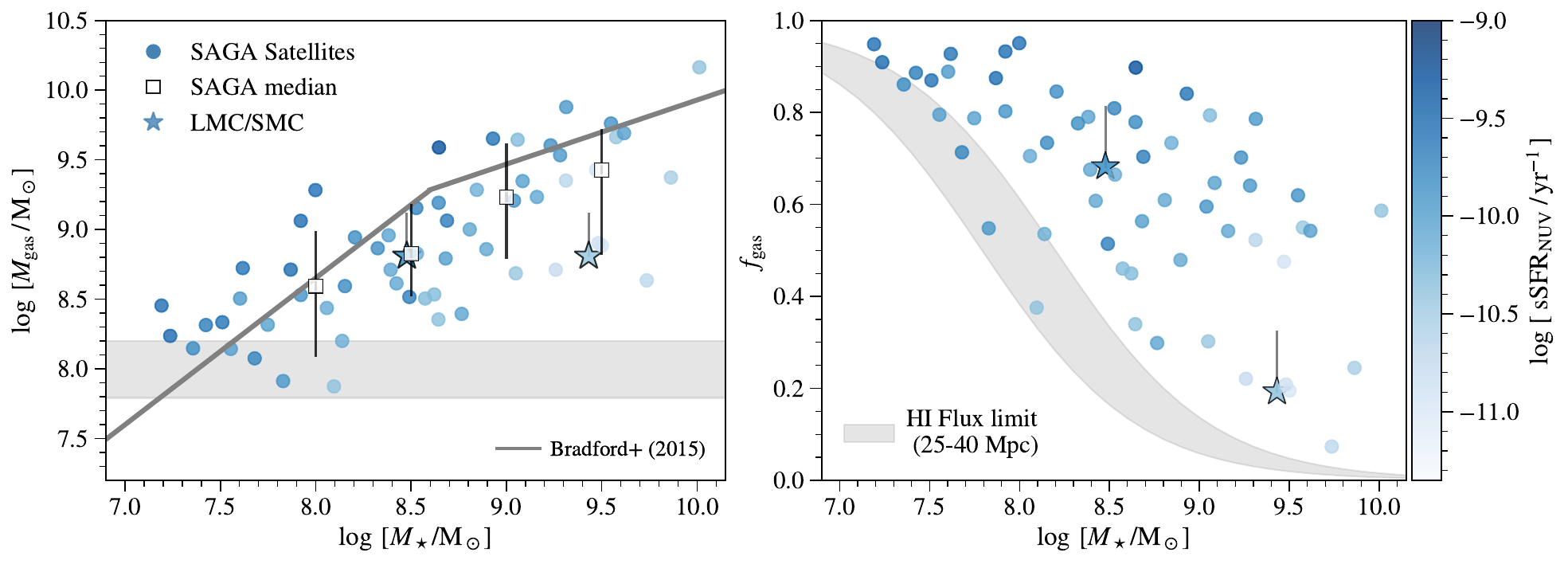}
    \caption{Literature measurements of atomic gas mass for 61 SAGA satellites:   gas mass versus stellar mass ({\bf left}) and gas fraction versus stellar mass ({\bf right}).   The LMC/SMC are shown as stars with small grey bars indicating the addition of MC Stream gas. In the left panel, we plot the empirical relationship from \citet{bradford2015} for isolated galaxies.  The median SAGA bins (white squares) fall below the Bradford et al.~relation.  In both panels, we compute the minimum gas fraction observable based on the ALFALFA HI flux limit of the ALFALFA for satellite between 25--40\,Mpc (grey region). In both panels, satellites are color-coded by sSFR$_{\rm NUV}$; gas-poor satellites have lower sSFR$_{\rm NUV}$. }
    \label{fig:HI}
\end{figure*}

In fitting the gas-phase metallicity-stellar mass relationship (the 'mass-metallicity' relationship), we include only satellites with S/N > 4 in both emission lines (Eq.\,6).   The [NII] line is weaker compared to H$\alpha$, even in our highest metallicity systems.  Thus our S/N threshold is equivalent to S/N$_{\rm [NII]}$ > 4.  Setting too strict a S/N$_{\rm [NII]}$ threshold will bias the sample towards higher metallicity systems, however, given the dependence on stellar mass, this should not affect satellites in our Gold sample.   In Figure~\ref{fig:metallicity}, we plot satellites with S/N$_{\rm [NII]}$ > 4 (circles).  For comparison only, we plot satellites with 3 < S/N$_{\rm [NII]}$ < 4 (triangles).  

To show the overall trend in SAGA satellites, we bin SAGA satellites into stellar mass bins and plot the medians as large squares in Figure~\ref{fig:metallicity}, with error bars denoting the 16th and 84th percentiles in each bin. 
The average scatter in each bin is 0.1\,dex. For comparison, we also plot the LMC/SMC as stars; both the LMC and SMC fall below the gas-phase mass-metallicity relation of the SAGA satellites which we discuss further in Section~\ref{ssec:MCs}.

We compare the SAGA satellite metallicities to two samples that include a wider range of galaxy environments:  the SDSS sample first presented by \citet{tremonti2004} limited to ($z<0.035$), and the SAGA background sample discussed in \S\ref{ssec_sfs_lit}.  As described above, we use the same N2-based metallicity calibration for all samples.  For the SDSS sample, we re-determine metallicities using the line fluxes measured by MPA-JHU \citep{Brinchmann2004}.  We restrict the SDSS sample to $z<0.035$. Following \citet{tremonti2004}, we show the SDSS results by plotting a running median and $1\sigma$ scatter as black solid/dashed lines in Figure~\ref{fig:metallicity}.  The median scatter for SDSS is $\sigma_{\rm 12+log(O/H)} = 0.06$.  This is slightly lower than the \citet{tremonti2004} value of 0.10, likely because we are restricting to a lower redshift range.   The masses and metallicities of the SAGA background galaxies are consist and nearly indistinguishable from the SDSS sample.

While neither the SDSS nor the SAGA background galaxies span the full stellar mass range of the SAGA satellites, extrapolating the medians of these samples to $<10^8~\msun$, the SAGA satellites lie at slightly higher gas-phase metallicities at all stellar mass.  The scatter in the SAGA satellites is $\sigma_{\rm 12+log(O/H)} = 0.1$, nearly twice that of the SDSS and SAGAbg sample.   Additional scatter above the mass-metallicity relation could be driven by tidal stripping of stellar mass during infall, or the lack of pristine infalling gas for satellites in a MW-like environment.  A more direct comparison requires a detailed understanding of the incompleteness and biases in both samples, which we defer to a future study.

Finally, the SAGA satellites in  Figure~\ref{fig:metallicity} are color-coded by sSFR$_{\rm NUV}$.   Comparing satellites above and below the binned median gas-phase metallicities (squares), we note an anti-correlation in the sense that satellites with higher sSFRs lie at lower gas-phase metallicities for a given stellar mass.   For satellites in the Gold sample, the difference in sSFR between satellites above and below median is $\Delta_{\rm \log(sSFR_{NUV})} = $0.3--0.5, in the sense that satellites below the median have higher sSFR$_{\rm NUV}$.  This anti-correlation between SFR and gas-phase metallicity at fixed stellar mass \citep[sometimes called the ``fundamental metallicity relation''; for a review, see][]{Maiolino2019} has been observed in many other galaxy samples \citep[e.g.,][]{Mannucci2010,yang2024}, and is often thought to result from the infall of metal-poor gas both diluting the ISM and triggering star formation. The existence of such a trend among SAGA satellites---which, as noted above, may not accrete as much pristine gas inside MW-like environments---may imply that gas infall alone cannot drive the observed relationship between SFR and metallicity.
We consider this further in the following section as we discuss the gas content of the SAGA satellites.

\subsection{HI Gas Content of SAGA Satellites}\label{ssec_gas}

Gas is the key raw material for star formation.  We next explore trends with HI gas content using  compiled literature HI measurements.  As detailed in \S\ref{ssec_hi_lit}, we focus on single-dish HI measurements for homogeneity and compute atomic gas masses directly from HI flux measurements using our SAGA host distances.  There are 61 SAGA satellites with HI measurements; the majority are in the Gold sample.  Because these HI measurements are from all-sky HI surveys, we are biased towards detecting the most gas-rich satellites.  The grey region in both panels of Figure~\ref{fig:HI} is the minimum detectable gas quantities for satellites in the SAGA distance range (25--40\,Mpc) and an HI flux limit equal to the faintest detection in our sample (0.3\,mJy \kms, roughly the flux limit of the ALFAFA survey).

In the left panel of Figure~\ref{fig:HI}, we plot the atomic gas mass versus stellar mass.    We plot the SAGA median values and 16/84th percentiles binned by stellar mass (white squares).   We plot the empirical relationship determined by \citet{bradford2015} for isolated galaxies in the same stellar mass regime (solid line).  The SAGA satellites fall below the empirical gas-stellar mass relationship for isolated galaxies.  This is consistent with a picture in which gas is being stripped and/or SAGA satellites are  not accreting as much gas as isolated systems.  We plot the position of the LMC/SMC in both panels; the MCs lie within the SAGA satellite distributions, and we discuss them further in \S\ref{ssec:MCs}.

In the right panel of Figure~\ref{fig:HI}, we plot the gas fraction, defined as $f_{\rm gas} = M_{\rm gas}/(M_{\rm gas} + \mstar)$, versus the stellar mass.  Given that we are biased towards detecting only the most gas-rich satellites, we see a trend in the sense that the maximum observed gas fraction increases towards lower stellar mass.  This is consistent with galaxies in a wider range of environments \citep{bradford2015}.

A strong trend is seen in both panels of Figure~\ref{fig:HI}  between the gas fraction and sSFR$_{\rm NUV}$ (color-coding).  Satellites with higher gas fractions show higher sSFRs (darker colors) at all stellar masses.  
That is, galaxies with more atomic gas have higher sSFRs, despite atomic gas not being the primary fuel for star formation.     This is likely related to the trend with metallicity in Figure~\ref{fig:metallicity}, which showed that satellites with higher sSFRs also have lower gas-phase metallicities for a given stellar mass.   These secondary trends between stellar mass, gas mass, metallicity and SFRs agree with ``gas-equilibrium'' models, in which galaxy properties are set by the balance between gas inflow and gas consumption: in environments where pristine gas is available, it is accreted onto a galaxy, leading to higher gas fractions, higher sSFR, and lower gas-phase metallicities.  At the same time, galaxies are consuming gas, driving down the gas fractions and increasing gas-phase metallicities \citep[e.g.,][]{torrey2019}.   This model primarily describes systems that are in equilibrium, and galaxies far from equilibrium (e.g., interacting pairs, barred galaxies) are not expected to obey these global relations \citep{Bustamante2020, Ellison2011}. It is therefore interesting to consider why many of the SAGA satellites do exhibit these trends, despite presumably interacting with their MW-like hosts. More detailed observations---especially of the outlier SAGA satellites that do not appear to follow these trends---are needed to further examine these relationships.

\begin{figure}[t!]
    \centering
    \includegraphics[width=1.01\columnwidth,clip,trim=5 5 0 0]{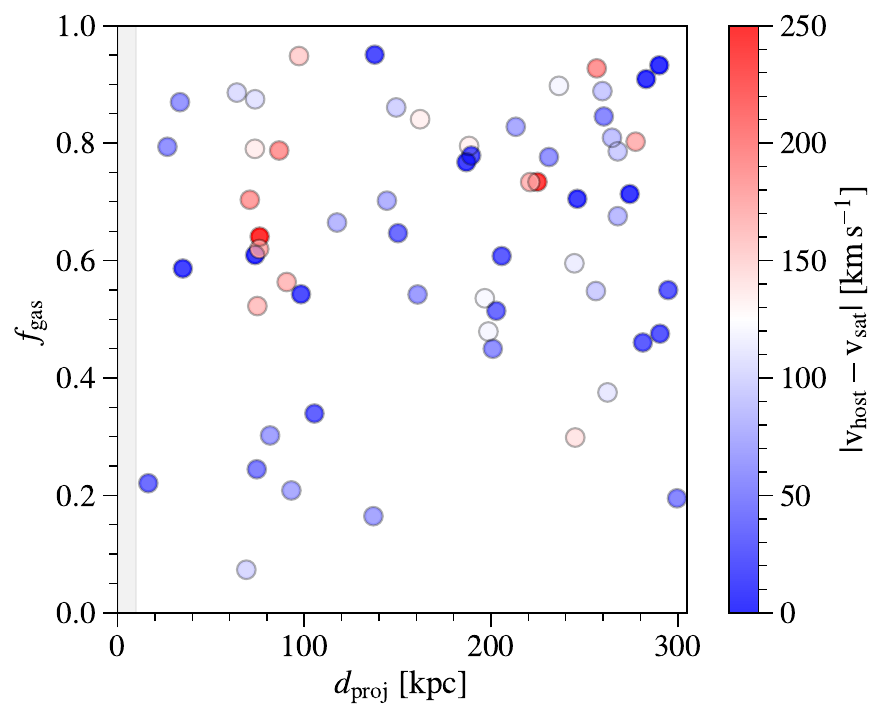}
    \caption{Gas fraction plotted against projected distance from host, color-coded by the relative velocity separation between satellite and host.   Satellites with low gas fractions are seen at all projected radii, but are more tightly bound  to their hosts (smaller relative velocities) as compared to satellites with higher gas fractions.}
    \label{fig:metal_resid}
\end{figure}

In Figure~\ref{fig:metal_resid}, we plot the gas fraction as a function of the projected distance from the host galaxy.  Gas rich satellites ($f_{\rm gas} > 0.5$) are distributed at all projected radii.   HI detection limits restrict gas poor satellites ($f_{\rm gas} < 0.5$) to only a handful of the most massive satellites.  While the majority of gas-poor systems lie inside of 100\,kpc, we do see a smaller number of gas poor satellites at large projected distance, in excess of what might be expected from interloper corrections alone.  \citet{jones2023b} found no gas poor satellites in SAGA DR2 hosts beyond 200\,kpc, however, this is likely due to small sample size.

We color-code satellites in Figure~\ref{fig:metal_resid} by the velocity difference between the satellite and its host.   Satellites which are more tightly gravitationally bound to its hosts will lie at closer projected separations and smaller (bluer) relative velocity differences.   Galaxies with the lowest gas fractions have smaller velocity differences relative to their host (< 150\,\kms) regardless of their projected separation.  Splitting our sample by gas fraction ($f_{\rm gas}=0.5$):   \edit {the mean relative velocity difference of high gas fraction satellites is $23\pm 14$\,\kms}~higher (less bound) as compared to low gas fraction satellites.   If we restrict the sample to satellites with stellar mass greater than $10^{8.5}\msun$, where HI flux sensitivity is highest, the difference between samples \edit{increases to $32\pm18$\,\kms.}   This hints that the lowest gas fraction galaxies (which trend towards lower sSFR and higher gas-phase metallicities) are also more gravitationally bound to their hosts.  Increasing the number of SAGA satellites with HI measurements is needed to further understand the roles of environment and star formation feedback in these systems.

\begin{figure*}[htb!]
    \centering
    \includegraphics[width=1.0\textwidth]{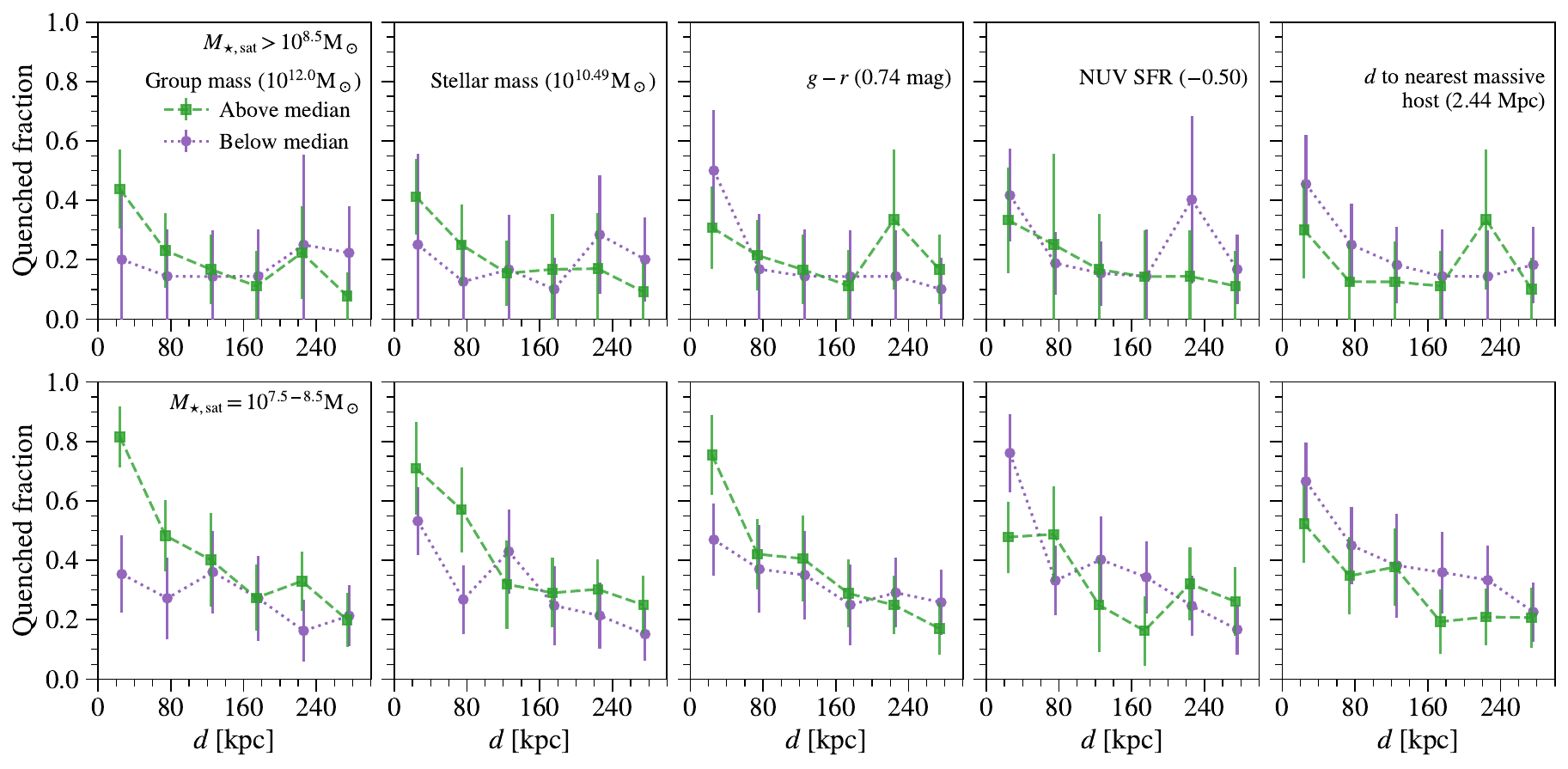}
    \caption{The quenched fraction of SAGA satellite versus projected distance from host. In all panels, we split the sample by host property, plotting the quenched fraction of systems above (green squares) and below (purple circles) the median value of ({\bf left to right}): Group mass (a proxy for halo mass), host stellar mass, host $g-r$ color, SFR$_{\rm NUV}$, and distance to nearest massive galaxy. The parenthetical numbers are the median values of these host properties. Because quenched fractions depend strongly on stellar mass, we plot separately satellites with stellar masses above $10^{8.5}\msun$ ({\bf top}) and between $10^{7.5-8.5}\msun$ ({\bf bottom}).}
    \label{fig:conform}
\end{figure*}

\section{Quenched Fractions and Host Properties}\label{ssec_conformity}

In this section, we return to the satellite quenched fractions discussed in \S\ref{sec:quenched} and investigate whether this quantity correlates with properties of the host galaxy.    Our analysis so far has implicitly assumed that host environments are identical across all 101 SAGA MW-mass systems. However, even though we impose stellar mass and environmental criteria on SAGA hosts (see Sec.~2 of \paperthree{}), those selections result in a range of host galaxy properties and environments.   Our host galaxies are selected over an absolute magnitude range $-23 > M_K > -24.6$, as a proxy for stellar mass, with no criteria on host SFRs: roughly half our hosts are star-forming as seen in Figure~\ref{fig:SFMS_hosts}; our isolation criteria allow some host galaxies to have MW-mass companions. 

Correlations between the quenched fraction of satellites and host properties, in particular the star-forming properties of the host, are related to studies of {\it galactic conformity}.   This term was first used by \citet{Weinmann2006} who found that both the color and SFR of satellite galaxies in SDSS depend on the color/SFR of a group’s central galaxy at fixed host halo mass (`one-halo' conformity).   Other studies extended this analysis to larger scales, finding host-satellite properties can be correlated beyond the virial radius out to distances 1--5\,Mpc \citep['two-halo' conformity; see e.g.,][]{kauffmann2013, hearin2015,treyer2018}.

In \paperthree{} we explored correlations between satellite abundance (number of satellites) and host/system properties, observing that the brightest satellite mass exhibits the strongest correlation with satellite abundance.   It is natural to expand this analysis to the fraction of quenched satellites in a system.  However, unlike the number of satellites, the quenched fractions depend on both stellar mass and projected distance to the host.  We split our SAGA host sample by various host properties, and directly examine the difference between quenched fractions as a function of stellar mass and projected distance. 

In Figure~\ref{fig:conform}, we plot the satellite quenched fraction as a function of projected radius from host, splitting our 101 SAGA hosts in half based on different host properties.   In each panel, we split systems by the median of various host properties.     We plot in Figure~\ref{fig:conform} a selection of properties (from left to right panels):  group mass (a proxy for halo mass inferred from group membership by \citealt{2017MNRAS.470.2982L}), stellar mass of the host, $g-r$ color of the host, SFR$_{\rm NUV}$ (determined in \S\ref{ssec_nuv_measure}), and distance to the nearest massive galaxy.   Because quenched fractions also vary with stellar mass (\S\ref{ssec:fq}), we  split these plots into two stellar mass bins (top and bottom panels). The top panels show only massive satellites with LMC-like masses ($ > 10^{8.5}\,\msun$), and the bottom panels show the SAGA Gold sample excluding the massive satellites.  

The most significant signal seen in Figure~\ref{fig:conform} is with group mass (halo mass of the SAGA host).    This halo mass estimation is based on both the sum of the luminosities of member galaxies and the difference in luminosity between the central and member galaxies, with the members identified by a group finder \citep{2017MNRAS.470.2982L}.  The quenched fractions are higher for systems with larger group halo mass, although this difference is only seen in the inner 100\,kpc.  \edit{ Similar to \paperthree, we calculate the Spearman’s rank correlation coefficient for the quenched fraction distribution as a function of (unbinned) host quantity.   For group halo mass, we find a trend of $\sim\,3\sigma$ significance ($p\text{-value} = 0.008$) if we restrict to within 100 kpc, but this trend is not significant when we include satellites throughout the virial volume ($p\text{-value} = 0.3$).  The difference is only significant for our lower stellar mass bin.   We also see a hint of this trend with host stellar mass, but at lower significance.   However, stellar mass is correlated with halo mass, so this may be a secondary correlation.  We see no significant trend with host color or NUV SFR.}   This suggests that higher halo mass SAGA systems are able to preferentially quench their innermost satellites.   One interpretation of this trend is that these systems host more massive hot halos which increase the efficiency of ram pressure stripping processes in the inner region.  \edit{ More generally, this suggests that the quenching mechanism associated with these host properties may start only after the satellites fall into the system.}

In the rightmost panel of Figure~\ref{fig:conform}, we split hosts based on the 
distance to a nearby neighbor.   This is an indicator of the density at a larger scale {\edit and simulations suggest that this may have an impact on satellite properties \citep[e.g.,][]{samuel2022, vannest2023}}.
Here a massive neighbor is defined as a galaxy with an absolute $K$-band magnitude brighter than $M_K = -23$, which is the lower-mass bound of the SAGA host selection. For each host, we determine the distance to the nearest galaxy with similar or larger stellar mass.   By definition, the minimum nearest-neighbor distance is 300\,kpc, the SAGA host virial radius.  For the MW itself, the nearest-neighbor distance is the distance to M31 (780\,kpc) although it might be smaller when observed in projection.   The median nearest-neighbor distance for the SAGA sample is 2.4\,Mpc.   
\edit{We observed a small difference in quenched fractions at all projected radii between systems with closer and further nearest massive galaxies; this difference is mainly driven by a handful of systems that have low quenched fraction and are also most isolated. However, if we calculate the Spearman’s rank correlation between quenched fractions and the distance to the nearest massive galaxy for the whole SAGA sample, the correlation is not statistically significant ($p$-value = 0.3).}

\section{Discussion} \label{sec:discussion}

\subsection{Connection to Theory}\label{ssec:theory}

Our goal in this paper is to understand the environmental processes that suppress or quench star formation in MW-like environments.    While our results provide qualitative constraints, a quantitative assessment of quenching mechanisms or timescales requires a broader modeling effort in a cosmological context.   In \paperfive{}, we interpret the SAGA results by extending the empirical galaxy--halo connection framework  \textsc{UniverseMachine} \citep{2019MNRAS.488.3143B,2021ApJ...915..116W} into the SAGA regime.   

In this section, we highlight key SAGA results which might best inform implementations of galaxy formation physics in semi-analytic models and hydrodynamical simulations.  A natural measure of a model's success is often a direct comparison to the MW satellites.  Yet the MW's quenched fraction as a function of stellar mass is more extreme than a typical SAGA host (\S\ref{ssec:fq});  as a function of projected radius (\S\ref{ssec:fq_rad}), the MW's quenched fraction is inverted relative to a typical SAGA host for stellar mass above $10^{7.5}\msun$.  Furthermore, the LMC/SMC themselves are not entirely typical of SAGA satellites at the same stellar mass (\S\ref{ssec:MCs}).   Surveys of satellites around the Milky Way and local volume have been used to assess quenching times and mechanisms \citep[e.g.,][]{10.1093/mnras/stv2058, 2021ApJ...909..139A,greene2023}.   The SAGA Survey significantly expands the available samples when assessing simulations of MW-like environments and beyond.

In the stellar mass range of the SAGA Survey, isolated field galaxies are predominantly star-forming \citep[e.g.,][]{geha2012,prole2021,Carleton2023}.  Naively assuming all infalling satellites are initially star-forming, the SAGA quenched fraction as both a function of stellar mass and projected radius provides strong constraint on quenching in a MW-like environment.    However, many complications exist including group pre-processing \citep{Roberts2017} and details of gas stripping during infall \citep{samuel2023,Zhu2024}.   Efficiency of stripping depends on satellites' orbital histories, properties of the host halo, and satellites' internal structure and is sensitive to the assumed sub-grid physics of star formation and supernova feedback.  Accurately modeling gas stripping thus requires high resolution hydrodynamic solvers \citep{Tonnesen2019}.  Additional constraints provided by SAGA include the observation that the lowest sSFR quartile of SAGA star-forming satellites follow a similar projected radial distribution as our quenched satellites (\S\ref{ssec_radial}) and secondary correlations between sSFR, atomic gas content and gas-phase metallicity (\S\ref{ssec_gas}).

While the SAGA Survey probes only a small slice of environments present throughout the Universe, MW-like regions lie between environmental extremes and can thus provide some constraint on quenching processes from the most isolated to dense environments.   In this context, reproducing host-to-host scatter within the SAGA sample (\S\ref{ssec_quenched}) and correlations across host properties (\S\ref{ssec_conformity}) ---which are expected to correlate with host halos' formation histories---is critical.   For example, the correlation between quenched fraction and the distance to a nearby massive galaxy neighbor hints at a large-scale environmental dependence of quenching/stripping \citep[e.g.,][]{vannest2023,Christensen2023}.  Finally, improved modeling of SAGA systems will allow us to more robustly interpret observations of the Milky Way itself, particularly given the unique formation history of the Milky Way's satellite system discussed below.

\subsection{The LMC/SMC in the Cosmological Context}\label{ssec:MCs}

A major goal of the SAGA Survey is to understand the MW system itself in a cosmological context.   In \paperthree{}, \S6,1, we concluded that although the MW differs in many respects from the average SAGA system (e.g., satellite abundance, radial concentration), these differences are reconciled if the MW is an older, slightly less massive host with a recently accreted LMC/SMC system.   In this section, we focus on the LMC/SMC themselves, asking how these recently accreted satellites compare to satellites with similar luminosities in the SAGA Survey.   


We first plot the LMC and SMC in Figure~\ref{fig:SFMS_hosts}.   We adopt stellar masses from \citet{vandermarel2009} of $2.7\times10^{9}$\,\msun~and $3.1\times10^{8}$\,\msun~(equivalent to $\log[M_\star/\msun] =$ 9.4 and 8.5) for the LMC and SMC, respectively.   We determine the SFR by averaging over the resolved star formation history of each \citep{Bolatto2011,Massana2022}, matched to the timescale of our SFR indicator.  In comparing to our measured NUV-based SFR, we average over the past 100\,Myr and adopt 0.1\,\msun\,yr$^{-1}$ and 0.04\,\msun\,yr$^{-1}$~for the LMC and SMC.  This corresponds to a specific star formation rate of $\log[(\text{sSFR}/\text{yr}^{-1})] = -10.4$ and $-9.9$.   

The optical colors of the LMC/SMC are bluer than any SAGA satellite with comparable stellar mass (see \paperthree{}, Figure 12, \citealt{Tollerud2011, robotham2012}).    \citet{Tollerud2011} proposed that the LMC/SMC's color may be due to  star formation triggered during infall into the MW environment.   We can rule this hypothesis out by comparing the current SFR of these satellites to the SAGA distribution at a similar stellar mass.  Plotting these values in  Figure~\ref{fig:SFMS_hosts}, the LMC and SMC have fairly normal SFR and sSFR compared to SAGA satellites at the same stellar mass.   From the median values provided in Table~\ref{table:log_sm}, the LMC has lower than median sSFR, while the SMC is higher, but both satellites are within the 16/84th percentile scatter of SAGA satellites at the same stellar mass.

An alternative explanation for the blue LMC/SMC colors could be metallicity \citep{pan2023}. The LMC and SMC are conventionally considered to have metallicities of $\sim50\%$ and $\sim20\%$ solar, respectively.  We place the LMC/SMC on the SAGA mass-metallicity relationship (Figure~\ref{fig:metallicity}) using direct temperature metallicity measurements of HII regions in the LMC and SMC, taking the gas-phase metallicities of their brightest star-forming regions, as is done when placing spectroscopic fibers on the SAGA satellites.    Direct  measurements from spectroscopy yield metallicities of 12+log(O/H)$\approx8.36$ for the LMC and 12+log(O/H)$\approx8.1$ for the SMC \citep{DominguezGuzman2022}.   These measurements indicate the LMC and SMC have relatively low metallicities compared to SAGA satellite and background galaxies at the same stellar mass, consistent with their blue optical colors. However, we caution that different methods can shift metallicity estimates of the LMC/SMC by $\sim\,0.2$ dex.

The gas content of the LMC/SMC, on the other hand, are well within the SAGA distribution.  We place the LMC/SMC on Figure~\ref{fig:HI} using the HI mass estimates of $4.6\times10^8\msun$ for both the LMC and SMC \citep{putman2021}.  While the LMC is below the median SAGA value, the SMC agrees with the median.   We also indicate the HI mass associated with the MC Stream \citep[$5\times10^8\msun$;][]{Bruns2005,Nidever2010}.  We indicate the Stream gas on both MCs since it could be stripped from either Clouds.   If the Stream originates entirely from the LMC, this would place the LMC well within the SAGA distribution.

Taken together, these results suggest that the LMC/SMC are not extreme outliers among the SAGA satellites, except perhaps in their relatively low metallicities (which produce bluer optical colors). As noted in \S~\ref{ssec_metallicity}, these low metallicities may also be due to the recent infall of the LMC/SMC system: as satellites interact with their hosts, tidal effects could remove stellar mass or prevent the accretion of pristine gas, leading to higher gas-phase metallicities over time.  This implies that the LMC/SMC are primarily unique due to their recent infall onto the MW, rather than any of their intrinsic properties.
More detailed observations of massive satellites in the SAGA sample---including whether these massive satellites have satellites of their own---are needed to confirm this.

\section{Summary} \label{sec:summary}

In this paper we examine quenched fractions, star formation rates, gas-phase metallcities and atomic gas masses for \nsats\ satellites around 101 Milky Way analogs in the SAGA Survey.   Our goal is to better understand the environmental processes that to suppress or quench star formation in MW-like environments.   We measure star-formation rates for the majority of SAGA satellites based on both H$\alpha$ EWs and NUV fluxes.  For a subsample of satellites, we measure additional line fluxes, examining the BPT diagram and measure gas-phase metallicities.   Our main results are:

\begin{enumerate}

\item   SAGA quenched fractions increase with decreasing stellar mass, from $15\pm5$\% above $10^{8.5}\msun$ (majority star-forming satellites) to $79\pm3$\% at $10^{7}\msun$ (majority quenched).
The Milky Way's quenched fraction is a 1-$\sigma$ outlier from the average SAGA quenched fraction distribution (\S\ref{ssec:fq}).

\item   Quenched satellites with $\mstar > 10^{7.5}\msun$ are more centrally concentrated than star-forming satellites.  The half-radius for quenched satellites is $92_{-26}^{+21}$\,kpc compared to $183_{-13}^{+7}$\,kpc for star-forming satellites (\S\ref{ssec:fq_rad}).

\item The SFR--stellar mass relationship of SAGA satellites is similar when SFR is measured in either H$\alpha$ or NUV, however, H$\alpha$ shows more scatter and is skewed towards higher SFRs (\S\ref{ssec_sfs}).

\item Within the star-forming population, the radial distribution of satellites in the lowest sSFR quartile is more centrally concentrated than higher sSFR satellites and is more similar to the quenched satellite population than the rest of the star-forming population.   The sSFR of the SAGA star-forming population increases with decreasing stellar mass and is roughly constant with projected radius~(\S\ref{ssec_radial}).   

\item We place a lower limit of 0.3\% on the AGN occupation fraction in SAGA satellites based on a single SAGA satellite with a central bright radio continuum source (\S\ref{ssec:bpt}).

\item The median gas-phase metallicity of SAGA satellites is higher than field galaxies and shows larger scatter (\S\ref{ssec_metallicity}).

\item The median atomic gas mass of SAGA satellites with literature HI measurements falls below the empirical gas mass-stellar mass relationship determined for isolated galaxies by \citet{bradford2015}.   Satellites with the lowest gas fractions are more tightly bound to their host (\S\ref{ssec_gas}).

\item At a given stellar mass, satellites with lower sSFR$_{\rm NUV}$ have lower than average gas fractions and higher gas-phase metallicity (\S\ref{ssec_metallicity}, \S\ref{ssec_gas}).

\item  Satellite quenched fractions are higher in systems with higher host halo mass, but higher quenched fractions are only seen in the inner 100\,kpc.  \edit{We do not see significant trends with host color or NUV SFR} (\S\ref{ssec_conformity}).

\end{enumerate}

Distilling these results further, the SAGA satellites are preferentially quenched towards lower stellar masses and smaller projected radii.  Quenched galaxies are more radially concentrated compared to the star-forming satellites.    Within the star-forming population, satellites in the lowest quartile of specific star formation are also more radially concentrated.   Satellites with low gas fractions appear more bound to their hosts compared to satellites with more gas.    These observations all point to quenching processes that are not instantaneous, but instead require a full orbit or more to remove gas and quench star formation.  

The SAGA Survey represents the largest sample of MW-like satellites down to our completeness limit of $\mstar \sim 10^{6.75}\msun$.   While we have explored many properties of these satellites, numerous further studies are possible with these data.  As detailed in \paperthree{}, \S6.3, planned future observations include higher spectral resolution follow-up to measure dynamical masses and improved metallicities for our known SAGA satellites.     To facilitate future work for other investigators, the measured properties for all satellites are made available in \satstable.

\medskip
The authors thank Jenny Greene, Ananthan Karunakaran,  Chris Lidman, Ragadeepika Pucha, Jenna Samuel, and Andrew Wetzel for helpful discussions and feedback that have improved this manuscript. 

M.G.\ and Y.A.\ were supported in part by a grant to M.G~from the Howard Hughes Medical Institute (HHMI) through the HHMI Professors Program.
Support for Y.-Y.M.\ during 2019--2022 was in part provided by NASA through the NASA Hubble Fellowship grant no.\ HST-HF2-51441.001 awarded by the Space Telescope Science Institute, which is operated by the Association of Universities for Research in Astronomy, Incorporated, under NASA contract NAS5-26555.

This research used data from the SAGA Survey (Satellites Around Galactic Analogs; sagasurvey.org). The SAGA Survey is a galaxy redshift survey with spectroscopic data obtained by the SAGA Survey team with the Anglo-Australian Telescope, MMT Observatory, Palomar Observatory, W. M. Keck Observatory, and the South African Astronomical Observatory (SAAO). The SAGA Survey also made use of many public data sets, including: imaging data from the Sloan Digital Sky Survey (SDSS), the Dark Energy Survey (DES), the GALEX Survey, and the Dark Energy Spectroscopic Instrument (DESI) Legacy Imaging Surveys, which includes the Dark Energy Camera Legacy Survey (DECaLS), the Beijing-Arizona Sky Survey (BASS), and the Mayall z-band Legacy Survey (MzLS); redshift catalogs from SDSS, DESI, the Galaxy And Mass Assembly (GAMA) Survey, the Prism Multi-object Survey (PRIMUS), the VIMOS Public Extragalactic Redshift Survey (VIPERS), the WiggleZ Dark Energy Survey (WiggleZ), the 2dF Galaxy Redshift Survey (2dFGRS), the HectoMAP Redshift Survey, the HETDEX Source Catalog, the 6dF Galaxy Survey (6dFGS), the Hectospec Cluster Survey (HeCS), the Australian Dark Energy Survey (OzDES), the 2-degree Field Lensing Survey (2dFLenS), and the Las Campanas Redshift Survey (LCRS); HI data from the Arecibo Legacy Fast ALFA Survey (ALFALFA), the FAST all sky HI Survey (FASHI), and HI Parkes All-Sky Survey (HIPASS); and compiled data from the NASA-Sloan Atlas (NSA), the Siena Galaxy Atlas (SGA), the HyperLeda database, and the Extragalactic Distance Database (EDD). The SAGA Survey was supported in part by NSF collaborative grants AST-1517148 and AST-1517422 and Heising–Simons Foundation grant 2019-1402. SAGA Survey’s full acknowledgments can be found at \https{sagasurvey.org/ack}.

\software{
    Numpy \citep{numpy, 2020NumPy-Array},
    SciPy \citep{scipy, 2020SciPy-NMeth},
    numexpr \citep{numexpr},
    Matplotlib \citep{matplotlib},
    IPython \citep{ipython},
    Jupyter \citep{jupyter},
    Astropy \citep{astropy},
    easyquery (\https{github.com/yymao/easyquery}),
    adstex (\https{github.com/yymao/adstex})
}

\bibliographystyle{aasjournal}
\bibliography{main}

\appendix
\counterwithin{figure}{section}
\counterwithin{table}{section}

\section{Data Tables}
\label{app:tables}

In this appendix we provide two supplement data tables. 

\begin{enumerate}
    \item \autoref{table:log_sm} lists the satellite quenched fractions and star formation rates as a function of stellar masses. These data are used in Figures~\ref{fig:fq}, \ref{fig:SFMS_internal} and \ref{fig:SF_MS}.
    \item \autoref{table:radial} lists the satellite quenched fractions and specific star formation rates as a function of the satellites' project distances to their respective hosts. These data are used in Figures~\ref{fig:fq_rad} and \ref{fig:sf_radial}. 
\end{enumerate}

\setlength{\belowdeluxetableskip}{-10pt}

\begin{deluxetable}{c|cc|ccc|ccc|ccc} 
\tablecaption{SAGA Satellite Quenched Fractions and Star Formation Rates as a function of Stellar Masses \label{table:log_sm}}
\tablehead{
  \multicolumn{3}{c}{} &
  \multicolumn{3}{c}{NUV SFR} &
   \multicolumn{3}{c}{NUV sSFR} &
  \multicolumn{3}{c}{H$\alpha$ SFR} \\
  \cline{4-12}
  \colhead{$\log(\mstar/\msun)$} &
  \colhead{$f_\text{quenched}$} &
  \colhead{$\sigma(f_\text{quenched})$} &
  \colhead{Median} &
  \colhead{16\%} &
  \colhead{84\%} &
    \colhead{Median} &
  \colhead{16\%} &
  \colhead{84\%} &
  \colhead{Median} &
  \colhead{16\%} &
  \colhead{84\%} 
}
\startdata
9.38 & 0.135 & 0.049 & $-0.88$ & $-1.25$ & $-0.44$ & $-10.26$ & $-10.78$ & $-9.88$ & $-0.75$ & $-1.19$ & $-0.36$ \\
8.62 & 0.160 & 0.053 & $-1.65$ & $-1.90$ & $-1.24$ & $-10.27$ & $-10.53$ & $-9.83$ & $-1.53$ & $-1.97$ & $-0.89$ \\
8.15 & 0.259 & 0.061 & $-1.94$ & $-2.29$ & $-1.57$ & $-10.09$ & $-10.42$ & $-9.76$ & $-2.16$ & $-2.50$ & $-1.33$ \\
7.83 & 0.335 & 0.063 & $-2.19$ & $-2.61$ & $-1.79$ & $-10.02$ & $-10.51$ & $-9.62$ & $-2.20$ & $-2.64$ & $-1.76$ \\
7.57 & 0.408 & 0.061 & $-2.40$ & $-2.70$ & $-2.03$ & $-9.97$ & $-10.29$ & $-9.61$ & $-2.59$ & $-2.90$ & $-1.62$ \\
7.31 & 0.562 & 0.056 & $-2.47$ & $-2.88$ & $-2.24$ & $-9.78$ & $-10.15$ & $-9.55$ & $-2.61$ & $-3.10$ & $-2.01$ \\
6.95 & 0.792 & 0.026 & $-2.84$ & $-3.06$ & $-2.54$ & $-9.79$ & $-10.01$ & $-9.47$ & $-3.00$ & $-3.34$ & $-2.60$ 
\enddata
\tablecomments{Binned data used in Figures~\ref{fig:fq},  \ref{fig:SFMS_internal} and \ref{fig:SF_MS}.  The quenched fraction ($f_\text{quenched}$) is calculated as $N_q/N$, where $N_q$ and $N$ are the numbers of quenched and all satellites in that bin. The errors on quenched fraction are binomial (Poisson) errors $\sigma = \sqrt{p'(1-p')/N}$, where $p'$ is calculated using Maximum a Posteriori $(N_q+1)/(N+2)$. The SFR values are shown in $\log[\text{SFR} / (\msun \text{yr}^{-1})]$, and the sSFR values are in $\log[(\text{sSFR}/\text{yr}^{-1})]$. }
\end{deluxetable}

\begin{deluxetable}{c|cc|ccc|ccc} 
\tablecaption{SAGA Satellite Quenched Fractions and Specific Star Formation Rates as a function of Projected Distances to Host \label{table:radial}}
\tablehead{
  \multicolumn{3}{c}{} &
  \multicolumn{3}{c}{Gold Sample H$\alpha$ sSFR}  &
  \multicolumn{3}{c}{Silver Sample H$\alpha$ sSFR} \\
  \cline{4-9}
  \colhead{$d_\text{proj}$ [kpc]} &
  \colhead{$f_\text{quenched}$} &
  \colhead{$\sigma(f_\text{quenched})$} &
  \colhead{Median} &
  \colhead{16\%} &
  \colhead{84\%} &
  \colhead{Median} &
  \colhead{16\%} &
  \colhead{84\%} 
           }
\startdata
\phn34.2 & 0.518 & 0.071 & $-10.17$ & $-10.48$ & $-\phn9.45$ & $-\phn9.85$ & $-10.26$ & $-\phn9.51$ \\
\phn82.5 & 0.252 & 0.067 & $-10.27$ & $-10.59$ & $-\phn9.60$ & $-10.05$ & $-10.33$ & $-\phn9.42$ \\
130.8 & 0.247 & 0.073 & $-10.18$ & $-10.49$ & $-\phn9.85$ & $-10.06$ & $-10.42$ & $-\phn9.78$ \\
179.2 & 0.166 & 0.064 & $-10.05$ & $-10.48$ & $-\phn9.42$ & $-\phn9.99$ & $-10.32$ & $-\phn9.59$ \\
227.5 & 0.205 & 0.065 & $-\phn9.96$ & $-10.35$ & $-\phn9.33$ & $-\phn9.89$ & $-10.21$ & $-\phn9.34$ \\
275.8 & 0.132 & 0.048 & $-\phn9.92$ & $-10.39$ & $-\phn9.39$ & $-\phn9.80$ & $-10.36$ & $-\phn9.36$ 
\enddata
\tablecomments{Binned data used in Figures~\ref{fig:fq_rad} and \ref{fig:sf_radial}. The quenched fraction ($f_\text{quenched}$) is calculated as $N_q/N$, where $N_q$ and $N$ are the numbers of quenched and all satellites in that bin. The errors on quenched fraction are binomial (Poisson) errors $\sigma = \sqrt{p'(1-p')/N}$, where $p'$ is calculated using Maximum a Posteriori $(N_q+1)/(N+2)$. The sSFR values are derived from H$\alpha$ measurements, and are shown in $\log[\text{sSFR} / (\text{yr}^{-1})]$.}
\end{deluxetable}

\end{document}